\documentclass[aps,prc,amsfonts,preprintnumbers,superscriptaddress,showpacs,nofootinbib]{revtex4}
\pdfoutput=1
\usepackage{graphics}
\usepackage{amsmath}
\usepackage{amssymb}
\usepackage{psfrag}
\usepackage{epsfig}
\usepackage{epsf}
\usepackage{float}
\usepackage{slashed}
\allowdisplaybreaks

\newcommand{\fet}[1]{\mbox{\boldmath $#1$}}

\newcommand{\beq}{\begin{equation}}
\newcommand{\eeq}{\end{equation}}
\newcommand{\beqa}{\begin{eqnarray}}
\newcommand{\eeqa}{\end{eqnarray}}
\newcommand{\nn}{\nonumber \\ }

\begin{document}

\title{Improved chiral nucleon-nucleon potential up to next-to-next-to-next-to-leading order}

\author{E.~Epelbaum}
\email[]{Email: evgeny.epelbaum@rub.de}
\affiliation{Institut f\"ur Theoretische Physik II, Ruhr-Universit\"at Bochum,
  D-44780 Bochum, Germany}
\author{H.~Krebs}
\email[]{Email: hermann.krebs@rub.de}
\affiliation{Institut f\"ur Theoretische Physik II, Ruhr-Universit\"at Bochum,
  D-44780 Bochum, Germany}
\author{U.-G.~Mei{\ss}ner}
\email[]{Email: meissner@hiskp.uni-bonn.de}
\affiliation{Helmholtz-Institut~f\"{u}r~Strahlen-~und~Kernphysik~and~Bethe~Center~for~Theoretical~Physics,
~Universit\"{a}t~Bonn,~D-53115~Bonn,~Germany}
\affiliation{Institut~f\"{u}r~Kernphysik,~Institute~for~Advanced~Simulation,
and J\"{u}lich~Center~for~Hadron~Physics, Forschungszentrum~J\"{u}lich,~D-52425~J\"{u}lich,~Germany}
\affiliation{JARA~-~High~Performance~Computing,~Forschungszentrum~J\"{u}lich,~D-52425~J\"{u}lich,~Germany}
\date{\today}

\begin{abstract}
We present improved nucleon-nucleon potentials derived in chiral effective field theory 
up to next-to-next-to-next-to-leading order. We argue that the nonlocal momentum-space 
regulator employed in the two-nucleon potentials of Refs.~\cite{Epelbaum:2004fk,Entem:2003ft} is not the most efficient choice, 
in particular since it affects the long-range part of the
interaction. We are able to significantly 
reduce finite-cutoff artefacts by using an appropriate regularization in coordinate space which  
maintains the analytic structure of the amplitude. The new potentials do not require the additional 
spectral function regularization employed in Ref.~\cite{Epelbaum:2004fk} to cut off the short-range components of the two-pion 
exchange and make use of the low-energy constants $c_i$ and $d_i$ determined from pion-nucleon 
scattering without any fine tuning. We discuss in detail the construction of the new potentials and convergence 
of the chiral expansion for two-nucleon observables. We also introduce a new procedure
   for estimating the theoretical uncertainty from the truncation of the chiral expansion
   that replaces previous reliance on cutoff variation. 
\end{abstract}

\pacs{13.75.Cs,21.30.-x}

\maketitle

\vspace{-0.2cm}

%%%%%%%%%%%%%%%%%%%%%%%%%%%%%%%%%%%%%%%%%%%%%%%%%%%%%%%%%%%%%%%%%%%%%%%%%%%%%%%%%
\section{Introduction}
\def\theequation{\arabic{section}.\arabic{equation}}
\label{sec:intro}

In the past decade, we have witnessed impressive progress in the field
of low-energy nuclear physics which is, to
a large extent, related to exciting theoretical developments. 
On the one hand, rapidly increasing computational resources and
improvements in algorithms make selected nuclear physics observables
amenable to numerical simulations in lattice QCD, see
Ref.~\cite{Beane:2010em} for a review article.  
On the other hand, considerable progress has been achieved towards a
quantitative description of nuclear forces within the framework of chiral effective field
theory (EFT) initiated in the pioneering work of Weinberg
\cite{Weinberg:1990rz}. This approach has been used to derive nuclear
forces, defined as kernels of the corresponding dynamical equations,   
order-by-order within the EFT expansion, see
Refs.~\cite{vanKolck:1994yi,Ordonez:1995rz} for the first quantitative
studies along these lines. 
The resulting scheme, based on solving the nuclear $A$-body problem
in a Hamiltonian framework with interactions between nucleons tied to
QCD via its symmetries, has been developed into a
major research field in computational few- and many-body physics and provides
nowadays a commonly accepted approach to  \emph{ab-initio} studies of
nuclear structure and reactions \cite{Epelbaum:2008ga,Vary:2014pya}. In addition to
offering a natural explanation for
the observed hierarchy of many-body forces, $V_{2N} \gg V_{3N} \gg
V_{4N} \ldots$, and allowing for the estimation of the theoretical
uncertainty, it is expected to shed light on the long-standing
three-nucleon force (3NF) problem, an old but still very current 
topic in nuclear physics
\cite{KalantarNayestanaki:2011wz,Hammer:2012id}. 
While effects of 3NFs in low-energy nuclear 
observables are expected to be smaller than the ones of
the nucleon-nucleon (NN) force, their inclusion is
mandatory at the level of accuracy of todays few- and
many-body {\it ab-initio} calculations. However, in spite of decades of effort, the
structure of the 3NF is not properly described by the
available phenomenological models 
\cite{KalantarNayestanaki:2011wz}. Given the very rich spin-momentum
structure of the 3NF \cite{Krebs:2013kha,Phillips:2013rsa,Epelbaum:2014sea}, scarcer database for
nucleon-deuteron scattering as compared to the NN system and
relatively high computational cost of solving the Faddeev equations, further progress in this fields
requires substantial input from theory. This provides a strong motivation
to study the 3NF within chiral EFT, and this topic is recognized  as
an important frontier in the field
\cite{Vary:2014pya,Machleidt:2010kb,Epelbaum:2013tta}. In particular,
the recently formed Low Energy Nuclear Physics International
Collaboration (LENPIC) \cite{LENPIC} intends to carry out detailed  {\it ab-initio} 
calculations of few- and many-nucleon systems in order to study
effects of the 3NF complete through fourth order in the chiral
expansion, i.e.~next-to-next-to-next-to-leading order (N$^3$LO).  

Clearly, looking for fine details of the 3NF, which itself is expected to
provide a small correction to the dominant NN force, requires that
the NN interaction is known with a sufficiently high 
accuracy and that one is able to carry out reliable estimation of the
theoretical uncertainty. While accurate N$^3$LO NN potentials have been 
available since ten years \cite{Epelbaum:2004fk,Entem:2003ft}, there are certain issues which
might become relevant at the accuracy level of the ongoing and planned
calculations. First of all, the potentials of Ref.~\cite{Epelbaum:2004fk}
employ an additional spectral function regularization (SFR)
\cite{Epelbaum:2003gr,Epelbaum:2003xx} in order to suppress an
unphysically strong
attraction caused by the very strong short-range components of the subleading
two-pion exchange. 
On the other hand, the available calculations of the 3NF at and 
beyond the N$^3$LO level employ the standard dimensional
regularization. Introducing the additional SFR on some of the 3NF
contributions such as, for example, the so-called ring diagrams,
appears to be a nontrivial task. Notice that the potential of Ref.~\cite{Entem:2003ft} avoids
the usage of the SFR, but probably at the cost of allowing for a variable
functional form of the regulator function for different terms in the
interaction, see Table F.2 in Ref.~\cite{Machleidt:2011zz}. Another issue concerns
the adopted values of certain pion-nucleon ($\pi N$) low-energy constants (LECs) such as
especially the $c_i$'s, which accompany the subleading vertices in the $\pi N$
effective Lagrangian. These LECs govern the strength of the two-pion
exchange NN potential and of the long- and intermediate-range 3NFs and should be taken
consistently with the $\pi N$ system. It is well known that some of
these LECs and, especially, the LEC $c_3$ receive significant contributions
associated with the intermediate $\Delta$ excitation of the nucleon
and appear to be numerically large \cite{Bernard:1996gq}. It was found in Ref.~\cite{Epelbaum:2004fk} that
the large empirical values of $c_3$ would result in generating
unphysical deeply bound states in the NN system, so that a reduced
value for this LEC has been used  in the N$^3$LO potential of Ref.~\cite{Epelbaum:2004fk}. 
In the N$^3$LO potential of Ref.~\cite{Entem:2003ft}, the 
LECs $c_{2,3,4}$ were actually tuned to improve the quality of the fit
which resulted, in particular, in the value of $c_4$ incompatible with
the available determinations from the $\pi N$ system. Perhaps most
importantly, the theoretical uncertainty of the calculations due to
truncation of the chiral expansion at a given order was so far at best
estimated by means of a residual cutoff dependence. As 
will be argued below,  such an approach does not allow for a reliable
quantification of the theoretical accuracy. 

All these issues clearly call for taking a fresh look at the NN system
in chiral EFT. In this work, we introduce a new generation of the
chiral N$^3$LO NN potentials which make use of a local regularization of
the pion-exchange contributions. The resulting potentials provide an
excellent description of low-energy NN scattering observables and the
deuteron properties and resolve all the issues mentioned above. In
addition, we propose a simple approach for estimating the
uncertainty due to truncation of the chiral expansion, which does not
rely on cutoff variation, and study in detail the convergence of the
chiral expansion for various NN observables.

Our paper is organized as follows. In section~\ref{sec:ChiExpansion},
we discuss the chiral expansion for the NN potential up to N$^3$LO. 
The new regularization scheme is introduced in section~\ref{sec:Regular},
while section~\ref{sec:PhaseShifts} describes our
fit procedure and results for the phase shifts. The cutoff dependence
of the obtained predictions is addressed in section~\ref{sec:CutDep} while
our results for the deuteron properties can be found in section~\ref{sec:Deuteron}. 
The theoretical uncertainty of our results is discussed in section~\ref{sec:Uncertainty}, 
where we also analyze the convergence of the chiral expansion for various NN scattering
observables. Finally, the main findings of our work are summarized in
section~\ref{sec:Summary}.

\section{Chiral expansion of the two-nucleon potential}
\def\theequation{\arabic{section}.\arabic{equation}}
\label{sec:ChiExpansion}

Chiral effective field theory  provides a well-defined perturbative expansion for pionic and single-baryon 
observables as well as for nuclear forces and current operators. In
the two-flavor sector we are interested in here, the expansion parameter, denoted as $Q$, is defined as 
\beq 
\label{ExpPar}
Q \in \left\{ \frac{p}{\Lambda_b}, \; \frac{M_\pi}{\Lambda_b} \right\}\,,
\eeq
where $p$ refers to magnitude of three momenta of external particles,
$M_\pi$ is the pion mass and  
$\Lambda_b$  is the breakdown scale of chiral EFT. In the Goldstone boson and single-nucleon sectors, 
this scale can be naturally expected to be of the order of the $\rho$-meson mass $M_\rho$.\footnote{In the nucleon  
sector, one may expect the breakdown scale of the chiral
expansion to be lower due to the appearance of the $\Delta (1232)$
resonance. The effects of the intermediate $\Delta$ excitations can 
by systematically taken into account by treating the $\Delta$ isobar
as an explicit degree of freedom.} 
It was also argued \cite{Manohar:1983md} that   
it cannot be larger than the chiral symmetry breaking scale $\Lambda_\chi = 4 \pi F_\pi$, with $F_\pi \simeq 92\,$MeV 
the pion decay constant. On the other hand, in the few-nucleon
sector, calculations are usually carried out employing a finite
ultraviolet cutoff \cite{Lepage:1997cs}, whose value is typically chosen 
of the order of $\Lambda \sim 500\,$MeV. Using soft values of the cutoff
may effectively reduce the breakdown scale in the actual calculations. 
It is, therefore, more appropriate to estimate the breakdown scale 
of nuclear chiral EFT in a more conservative way rather than 
by $M_\rho \simeq 770\,$MeV or even 
the chiral symmetry breaking scale $\Lambda_\chi \simeq 1.2\,$GeV.  We
will discuss this issue in more detail in sections \ref{sec:CutDep}
and \ref{sec:Uncertainty}. 

\begin{figure}[tb]
\vskip 1 true cm
\includegraphics[width=\textwidth,keepaspectratio,angle=0,clip]{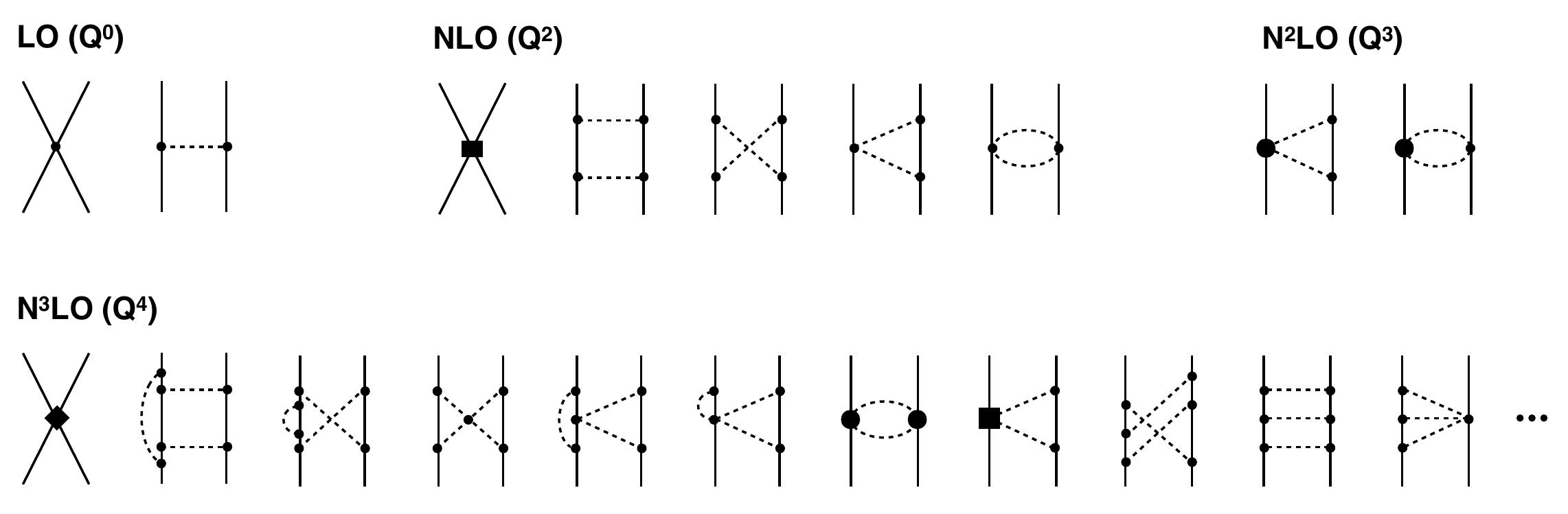}
    \caption{Chiral expansion of the NN potential  up to
      N$^3$LO. Solid and dashed lines refer to nucleons and pions,
      respectively. Solid dots denote the vertices from
      the lowest-order effective Lagrangians
      $\mathcal{L}_{\pi}^{(0)}$,   $\mathcal{L}_{\pi N}^{(0)}$ and
      $\mathcal{L}_{NN}^{(0)}$ with the superscript referring to the
      chiral dimension, for which we use the notation of
      Ref.~\cite{Epelbaum:2008ga}, see this work for explicit expressions.  
%see e.g.~Ref.~\cite{Epelbaum:2008ga} for more details and
%      explicit expressions. 
Filled
      circles refer to vertices from $\mathcal{L}_{\pi N}^{(1)}$ while
      squares (diamonds) denote vertices from $\mathcal{L}_{\pi
        N}^{(2)}$ and $\mathcal{L}_{N
        N}^{(2)}$ ($\mathcal{L}_{N
        N}^{(4)}$). Only those diagrams are shown which lead to
      contributions to the potential beyond renormalization of various
      coupling constants. Only irreducible contributions of various diagrams
      are taken into account in the potential as explained in the text.
\label{fig:graphs} 
 }
\end{figure}

Up to N$^3$LO, the NN potential involves contributions from one-, two- and three-pion exchange and 
contact terms with up to four derivatives which parametrize short-range interactions  
\beq
V = V_{\rm 1\pi} + V_{\rm 2\pi} + V_{\rm 3\pi} + V_{\rm cont} \,,
\eeq
see Fig.~\ref{fig:graphs} where the corresponding diagrams are shown. 
Here and in what follows, we adopt the standard power counting rules for short-range operators which are based 
on naive dimensional analysis, see Refs.~\cite{Nogga:2005hy,Long:2011xw,Birse:2010fj} for alternative suggestions 
and Refs.~\cite{Lepage:1997cs,Epelbaum:2006pt,Epelbaum:2009sd,Epelbaum:2012ua} for a related 
discussion. We further emphasize that diagrams shown in
Fig.~\ref{fig:graphs} do actually not correspond to Feynman graphs
which provide a graphical representation of the on-shell scattering
amplitude. Rather, they should be understood as a schematic
visualization of the irreducible parts of the amplitude, i.e.~those
diagrams which do not correspond to iterations of the dynamical
equation. For a comprehensive discussion on the various ways to derive
energy-independent nuclear
potentials and the associated unitary ambiguities see
Refs.~\cite{Kaiser:1997mw,Epelbaum:1998ka,Epelbaum:2005pn,Epelbaum:2007us,Epelbaum:2010nr,Machleidt:2011zz}.   

The static one-pion exchange potential (OPEP) is well known and takes the form 
\beqa
\label{ope_iso}
V_{1\pi}^{pp} &=& V_{1\pi}^{nn} = V_{1\pi} (M_{\pi^0}) \,, \nn
V_{1\pi}^{np} &=& -V_{1\pi} (M_{\pi^0}) + 2 (-1)^{I+1} V_{1\pi} (M_{\pi^\pm}) \,,
\eeqa
where $I$ denotes the total isospin of the two-nucleon system and 
\beq
V_{1\pi} (M_{\pi}) = - \frac{g_A^2}{4 F_\pi^2} \frac{\vec \sigma_1 \cdot
  \vec q \, \vec \sigma_2 \cdot \vec q}{q^2 + M_\pi^2}\,.
\eeq
Here and in what follows, $\vec{q} = \vec{p}\,'-\vec{p}$ refers to the
momentum transfer with $\vec{p}$ and $\vec{p}\, '$ 
being the  initial and final nucleon momenta in the  center--of--mass
system (cms), while $q \equiv | \vec q \, | $. 
Further, $\vec \sigma_i$ denotes the Pauli spin matrix of  nucleon $i$. Finally, $g_A$, $F_\pi$ and $M_{\pi^0}/M_{\pi^\pm}$
denote the axial-vector coupling constant of the nucleon, pion decay constant and neutral/charged pion mass, respectively. 
Notice that the above expressions include the isospin-breaking (IB) correction due to the different pion masses 
which is known to be the strongest long-range IB contribution, see 
Refs.~\cite{vanKolck:1997fu,Friar:1999zr,Walzl:2000cx,Friar:2003yv,Epelbaum:2005fd,Kaiser:2006ck} 
for more details on  the isospin dependence of 
the NN force. Charge dependence of the 
pion-nucleon coupling constant is consistent with zero
\cite{deSwart:1997ep} and for this reason will not be taken into
account in the present work. The form of the longest-range NN force specified 
above coincides with the one employed in the Nijmegen partial wave
analysis (NPWA) \cite{Stoks:1993tb} which we use as input 
for tuning the short-range  interactions. Relativistic corrections to the OPEP will be discussed at the end of this section. 

The chiral expansion of the two-pion exchange potential (TPEP) starts at
next-to-leading order (NLO) which corresponds to the chiral order $Q^2$. 
Using
the decomposition of the momentum-space TPEP 
\beqa
\label{2PEdec}
V_{2 \pi}  &=& V_C + \fet \tau_1 \cdot \fet \tau_2 \, W_C + \left[   
V_S + \fet \tau_1 \cdot \fet \tau_2 \, W_S \right] \, \vec \sigma_1 \cdot \vec \sigma_2 
+ \left[ V_T + \fet \tau_1 \cdot \fet \tau_2 \, W_T \right] 
\, \vec \sigma_1 \cdot \vec q \, \vec \sigma_2 \cdot \vec q \nn
&+& \left[ V_{LS} + \fet \tau_1 \cdot \fet \tau_2 \, W_{LS} \right] 
\, i (\vec \sigma_1 + \vec \sigma_2 ) \cdot ( \vec q \times \vec k ) \,,
\eeqa
where $\vec k = (\vec p + \vec p \, ')/2$, $\fet \tau_i$ denote the isospin Pauli matrices associated with the nucleon $i$, while 
$V_{C, S, T, LS}$ and   $W_{C, S, T, LS}$ are scalar functions which depend on the nucleon momenta,  
the order-$Q^2$ contributions take the form 
\beqa
\label{2PE_nlo}
W_C^{(2)} &=& - \frac{L (q )}{384 \pi^2 F_\pi^4}
\bigg[4M_\pi^2 (5g_A^4 - 4g_A^2 -1)  + q^2(23g_A^4 - 10g_A^2 -1)
+ \frac{48 g_A^4 M_\pi^4}{4 M_\pi^2 + q^2} \bigg] \,, \nn
V_T^{(2)} &=& -\frac{1}{q^2} V_S^{(2)}  = - \frac{3 g_A^4}{64
  \pi^2 F_\pi^4} \,L (q)\,,\nn
V_C^{(2)} &=& V_{LS}^{(2)} = W_S^{(2)} = W_T^{(2)} = W_{LS}^{(2)} = 0\,.
\eeqa
The loop function $L ( q)$ is defined in dimensional regularization (DR) via
\beq
\label{def_LA}
L ( q) = \frac{\sqrt{4 M_\pi^2 + q^2}}{q} \ln
\frac{\sqrt{4 M_\pi^2 + q^2} + q}{2 M_\pi} \,.
\eeq
Notice that we only list here non-polynomial in momenta contributions
while all polynomial 
terms are absorbed into contact interactions which will be discussed below.  

The corrections at order $Q^3$ giving rise to the subleading TPEP have the form 
\beqa
\label{2PE_nnlo}
V_C^{(3)}  &=& -\frac{3g_A^2}{16\pi F_\pi^4}  \bigg[2M_\pi^2(2c_1 -c_3) -c_3 q^2 \bigg] 
(2M_\pi^2+q^2) A ( q  )\,, \nn
W_T^{(3)} &=& -\frac{1}{q^2} W_S^{(3)}  = - \frac{g_A^2}{32\pi F_\pi^4} \,  c_4 (4M_\pi^2 + q^2) 
A( q )\,, \nn
V_S^{(3)} &=& V_{T}^{(3)} = V_{LS}^{(3)} = W_C^{(3)} = W_{LS}^{(3)} = 0\,,
\eeqa
where $c_{i}$ are LECs associated with the subleading $\pi \pi
NN$  vertices from $\mathcal{L}_{\pi N}^{(2)}$  and the loop function $A ( q )$ is given in DR by   
\beq
A ( q  ) = \frac{1}{2  q } \arctan \frac{ q }{2 M_\pi} \,.
\eeq 

At order $Q^4$, i.e. N$^3$LO, one encounters further corrections to the TPEP emerging from the various one- and two-loop diagrams
which have been calculated in Ref.~\cite{Kaiser:2001pc}. The contributions of the one-loop ``bubble'' diagrams to the TPEP 
take a particularly simple form 
\beqa
\label{2PE_nnnlo}
V_C^{(4)} &=& \frac{3}{16 \pi^2  F_\pi^4} \, L (q) \, 
\left\{ \left[ \frac{c_2}{6} (4M_\pi^2 + q^2) + c_3 (2 M_\pi^2 + q^2 ) - 4 c_1 M_\pi^2 
\right]^2 + \frac{c_2^2}{45} (4M_\pi^2 + q^2)^2 \right\} 
\nonumber \\
W_T^{(4)} &=& -\frac{1}{q^2} W_S^{(4)}= \frac{c_4^2}{96 \pi^2 F_\pi^4} (4M_\pi^2 + q^2)  
\, L(q)\,.
\eeqa
The remaining contributions from one- and two-loop diagrams can be most conveniently written 
using the (subtracted) spectral representation of the TPEP
\beqa
V_{C,S} (q) &=& - \frac{2 q^6}{\pi} \int_{2M_\pi}^\infty \, d \mu 
\frac{\rho_{C,S} (\mu )}{\mu^5 ( \mu^2 + q^2 )}\,, \quad \quad
V_T (q) = \frac{2 q^4}{\pi} \int_{2M_\pi}^\infty \, d \mu 
\frac{\rho_{T} (\mu )}{\mu^3 ( \mu^2 + q^2 )}\,, \nn
W_{C,S} (q) &=& - \frac{2 q^6}{\pi} \int_{2M_\pi}^\infty \, d \mu 
\frac{\eta_{C,S} (\mu )}{\mu^5 ( \mu^2 + q^2 )}\,, \quad \quad
W_T (q) = \frac{2 q^4}{\pi} \int_{2M_\pi}^\infty \, d \mu 
\frac{\eta_{T} (\mu )}{\mu^3 ( \mu^2 + q^2 )}\,, 
\eeqa
where $\rho_i$ and $\eta_i$ denote the corresponding spectral functions which are related to the 
potential via $\rho_i (\mu) = {\rm Im} V_i (i \mu )$,  $\eta_i (\mu) = {\rm Im} W_i (i \mu )$.   
For the spectral functions $\rho_i (\mu)$ ($\eta_i (\mu)$) one finds
\cite{Kaiser:2001pc}:
\beqa
\label{TPE2loop}
\rho_C^{(4)} (\mu ) &=& - \frac{3 g_A^4 (\mu^2 - 2 M_\pi^2 )}{\pi \mu (4 F_\pi)^6}
\, 
\bigg\{ (M_\pi^2 - 2 \mu^2 ) \bigg[ 2 M_\pi + \frac{2 M_\pi^2 - \mu^2}{2 \mu} 
\ln \frac{\mu + 2 M_\pi}{\mu - 2 M_\pi } \bigg] + 4 g_A^2 M_\pi (2 M_\pi^2 - \mu^2 )
\bigg\}\,, \nn
\eta_S^{(4)} (\mu ) &=& \mu^2 \eta_T^{(4)} (\mu ) = - 
\frac{g_A^4 (\mu^2 - 4 M_\pi^2 )}{\pi (4 F_\pi)^6}
\,  
\left\{ \left(M_\pi^2 - \frac{\mu^2}{4} \right) \ln \frac{\mu +  2 M_\pi}{\mu - 2 M_\pi }
+ (1 + 2 g_A^2 ) \mu M_\pi \right\}\,, \nn
\rho_S^{(4)} (\mu ) &=& \mu^2 \rho_T^{(4)} (\mu ) = - 
  \left\{
\frac{g_A^2 r^3 \mu}{8 F_\pi^4 \pi} 
(\bar d_{14} - \bar d_{15} ) - 
\frac{2 g_A^6 \mu r^3}{(8 \pi F_\pi^2)^3} \left[ \frac{1}{9} - J_1  + J_2 \right] \right\}\,, \nn
 \eta_C^{(4)} (\mu ) &=& \Bigg\{
\frac{r t^2}{24 F_\pi^4 \mu \pi} \left[ 2 (g_A^2 - 1) r^2 - 3 g_A^2 t^2 \right] (\bar d_1 + \bar d_2 ) \nn
&& {}+ \frac{r^3}{60 F_\pi^4 \mu \pi} \left[ 6 (g_A^2 - 1) r^2 - 5 g_A^2 t^2 \right] \bar d_3
- \frac{r M_\pi^2}{6 F_\pi^4 \mu \pi} \left[ 2 (g_A^2 - 1) r^2 - 3 g_A^2 t^2 \right] \bar d_5 \nn
&& {} - \frac{1}{92160 F_\pi^6 \mu^2 \pi^3} \Big[ - 320 (1 + 2 g_A^2 )^2 M_\pi^6 + 
240 (1 + 6 g_A^2 + 8 g_A^4 ) M_\pi^4 \mu^2 \nn
&& {}   \mbox{\hskip 3 true cm} - 60 g_A^2 (8 + 15 g_A^2 ) M_\pi^2  \mu^4
+ (-4 + 29 g_A^2 + 122 g_A^4 + 3 g_A^6 ) \mu^6 \Big] \ln \frac{2 r + \mu}{2 M_\pi} \nn
&& {} - \frac{r}{2700 \mu ( 8 \pi F_\pi^2 )^3} \Big[ -16 ( 171 + 2 g_A^2 ( 1 + g_A^2) 
(327 + 49 g_A^2)) M_\pi^4 \nn
&& {}   \mbox{\hskip 3 true cm} + 4 (-73 + 1748 g_A^2 + 2549 g_A^4 + 726 g_A^6 ) M_\pi^2 \mu^2 \nn
&& {}   \mbox{\hskip 3 true cm} - (- 64 + 389 g_A^2 + 1782 g_A^4 + 1093 g_A^6 ) \mu^4  \Big] \nn
&& {} + \frac{2 r}{3 \mu ( 8 \pi F_\pi^2 )^3} \Big[ 
g_A^6 t^4 J_1 - 2 g_A^4 (2 g_A^2 -1 ) r^2 t^2 J_2 \Big] \Bigg\}\,,
\eeqa
where we have introduced the abbreviations
\beq
r = \frac{1}{2} \sqrt{ \mu^2  - 4 M_\pi^2}\,, \quad \quad \quad
t= \sqrt{\mu^2 - 2 M_\pi^2}\,,
\eeq
and 
\beqa
J_1 &=& \int_0^1 \, dx \, \bigg\{ \frac{M_\pi^2}{r^2 x^2} - \bigg( 1 + \frac{M_\pi^2 }{r^2 x^2} \bigg)^{3/2}
\ln \frac{ r x + \sqrt{ M_\pi^2 + r^2 x^2}}{M_\pi} \bigg\}\,,\nonumber \\
J_2 &=& \int_0^1 \, dx \, x^2 \bigg\{ \frac{M_\pi^2}{r^2 x^2} - \bigg( 1 + \frac{M_\pi^2 }{r^2 x^2} \bigg)^{3/2}
\ln \frac{ r x + \sqrt{ M_\pi^2 + r^2 x^2}}{M_\pi} \bigg\}\,.
\eeqa
Here and in what follows, we use the scale--independent LECs $\bar d_{1}, \; \bar d_{2}, \; \bar d_{3}, \; \bar d_{5}, \; \bar d_{14}$ and 
$\bar d_{15}$ defined in \cite{Fettes:1998ud}. One also has to account for relativistic corrections to the TPEP which will be discussed 
at the end of this section. 

The short-range part of the chiral potential involves in the isospin
limit two derivative-less interactions contributing at leading order (LO), 
seven terms involving two derivatives at next-to-leading order (NLO)
and fifteen terms 
involving four derivatives at N$^3$LO. For isospin-breaking contact interactions, we employ here only the leading derivative-less 
terms which give rise to charge-independence and charge symmetry breaking in the $^1S_0$ NN phase shift. 
Notice that at N$^3$LO, one, strictly speaking, also needs to take
into account IB TPEP as well as the the $\pi \gamma$-exchange
potential.   
Given that we use the NPWA rather than experimental data 
as input to determine various LECs accompanying short-range interactions, we
employ here the same treatment of IB effects as used by the Nijmegen
group. Specifically, the only sources of 
(finite-range) IB contributions to the NN force are given by the OPEP, see Eq.~(\ref{ope_iso}) and the two derivative-less 
IB contact interactions. This is also exactly the same procedure as the one employed in our analysis reported in Ref.~\cite{Epelbaum:2004fk}.
The contact interactions used in the present work yield the following contributions to the 
NN potential in the partial-wave basis 
\beqa\label{VC}
\langle ^1S_0, \; {\rm np} | V_{\rm cont}^{\rm np}| ^1S_0, \;  {\rm np} \rangle&=& \tilde C_{1S0}^{\rm np} + C_{1S0} ( p^2 + p '^2) +
D_{1S0}^1 \, p^2 \, {p'}^2 + D_{1S0}^2 \, ({p}^4+{p}'^4)~,\nn
\langle ^1S_0, \;  {\rm pp} | V_{\rm cont}^{\rm pp}| ^1S_0, \;  {\rm pp} \rangle&=& \tilde C_{1S0}^{\rm pp}  + C_{1S0} ( p^2 + p '^2) +
D_{1S0}^1 \, p^2 \, {p'}^2 + D_{1S0}^2 \, ({p}^4+{p}'^4)~,\nn
\langle ^1S_0, \;  {\rm nn} | V_{\rm cont}^{\rm nn}| ^1S_0 , \;  {\rm nn} \rangle&=& \tilde C_{1S0}^{\rm nn}  + C_{1S0} ( p^2 + p '^2) +
D_{1S0}^1 \, p^2 \, {p'}^2 + D_{1S0}^2 \, ({p}^4+{p}'^4)~,\nn
\langle ^3S_1 | V_{\rm cont}| ^3S_1 \rangle&=& \tilde C_{3S1} + C_{3S1} ( p^2 + p '^2) +
D_{3S1}^1 \, p^2 \, {p'}^2 + D_{3S1}^2 \, ({p}^4+{p}'^4)~,\nn
\langle ^1P_1 | V_{\rm cont}| ^1P_1 \rangle&=& C_{1P1} \, p \, p' +
D_{1P1} \,p \, p' \, ({p}^2 +  p' \, ^2)~,\nn
\langle ^3P_1 | V_{\rm cont}| ^3P_1 \rangle&=& C_{3P1} \, p  \, p' + D_{3P1} \,p \, p' \, ({p}^2 \, 
 +p ' \, ^2 )~,\nn
\langle ^3P_0 | V_{\rm cont}| ^3P_0 \rangle&=& C_{3P0} \, p  \, p' + D_{3P0} \,p \, p' \, ({p}^2  
 +p ' \, ^2 )~,\nn
\langle ^3P_2 | V_{\rm cont}| ^3P_2 \rangle&=& C_{3P2} \, p  \, p' + D_{3P2} \,p \, p' \, ({p}^2  
 +p ' \,^2 )~,\nn
\langle ^1D_2 | V_{\rm cont}| ^1D_2 \rangle&=& D_{1D2} \, {p}^2\,{p}'^2~,\nn
\langle ^3D_2 | V_{\rm cont}| ^3D_2 \rangle&=& D_{3D2} \, {p}^2\,{p}'^2~,\nn
\langle ^3D_1 | V_{\rm cont}| ^3D_1 \rangle&=& D_{3D1} \, {p}^2\,{p}'^2~,\nn
\langle ^3D_3 | V_{\rm cont}| ^3D_3 \rangle&=& D_{3D3} \, {p}^2\,{p}'^2~,\nn
\langle ^3S_1 | V_{\rm cont}| ^3D_1 \rangle&=& C_{3D1 - 3S1} \, p^2 
+ D_{3D1 - 3S1}^1 \, {p}^2\,{p}'^2 + D_{3D1 - 3S1}^2 \, p^4~,\nn
\langle ^3D_1 | V_{\rm cont}| ^3S_1 \rangle&=& C_{3D1 - 3S1} \, {p '}^2 
+ D_{3D1 - 3S1}^1 \, {p}^2\,{p}'^2 + D_{3D1 - 3S1}^2 \, {p '}^4~,\nn
\langle ^3P_2 | V_{\rm cont}| ^3F_2 \rangle&=& D_{3F2 - 3P2}\, {p}^3\,{p}'~, \nn
\langle ^3F_2 | V_{\rm cont}| ^3P_2 \rangle&=& D_{3F2 - 3P2}\, {p}\,{p '}^3~, 
\label{V4ct}
\eeqa
where $\tilde C_{i}$, $\tilde C_{i}^{\rm np}$, $\tilde C_{i}^{\rm pp}$, $\tilde C_{i}^{\rm nn}$,
$C_{i}$ and $D_{i}$ denote the corresponding LECs. The relation between these LECs and the 
ones corresponding to the operator form of the short-range potential
can be found in Eq.~(2.6) of Ref.~\cite{Epelbaum:2004fk}.  Notice further that we do not show explicitly 
the pion mass dependence of various contact interactions. 

Finally, let us discuss the relativistic corrections to the potential
which according to our power counting scheme start to 
contribute at N$^3$LO. Notice that following
Ref.~\cite{Weinberg:1991um}, we treat the nucleon mass as a heavier
scale as compared with the  
breakdown scale $\Lambda_b$ by counting $Q/m_N \sim
Q^2/\Lambda_b^2$. 
The relativistic corrections are scheme-dependent or, more precisely, depend on the 
employed form of the dynamical equation and the choice of unitary transformations as explained in 
detail in Ref.~\cite{Friar:1999sj}. Contrary to the static N$^3$LO contributions, 
the results for the $1/m_N^2$-corrections to the OPEP and $1/m_N$-corrections to the  TPEP are not 
uniquely determined by the renormalizability requirement of the nuclear forces \cite{Bernard:2011zr} and depend on  
two arbitrary parameters, called $\bar \beta_{8,9}$ in that work, which correspond to the unitary ambiguity of the potential
and are related to the parameters $\mu$ and $\nu$ of   Ref.~\cite{Friar:1999sj}
via
\beq
\mu = 4 \bar \beta_9 + 1, \quad \quad \nu = 2 \bar \beta_8\,.
\eeq
Here and in what follows, we employ the ``minimal nonlocality'' choice of the potential corresponding to 
setting $\mu = 0$ and $\nu = 1/2$ or, equivalently, $\bar \beta_8 = 1/4$, $\bar \beta_9 = -1/4$.\footnote{We correct here 
a misprint in Eq.~(4.24) of Ref.~\cite{Bernard:2011zr}, where the ``minimal nonlocality'' choice was specified 
by  $\bar \beta_8 = 1/4$, $\bar \beta_9 = 0$.} This particular choice implies that the only $1/m_N^2$-corrections to the OPEP 
stem from accounting for the relativistic normalization of the nucleon field operators. Using the 
relativistic Schr\"odinger equation for NN scattering of the  form 
\beq
\label{schroed_rel}
\left[  2 \sqrt{p^2 + m_N^2}  + V \right] \Psi
=  2 \sqrt{k^2 + m_N^2} \Psi\,,
\eeq
where $m_N = m_p$ for the proton-proton (pp),  $m_N = m_n$ for the neutron-neutron
(nn)  and 
$m_N = 2 m_p m_n/(m_p + m_n)$ for the neutron-proton (np) case and $k$ denotes the
momentum corresponding to the energy eigenvalue $E$, the full non-static expression for the OPEP takes the form 
\beq
\label{1pep_rel}
V_{1 \pi} = \frac{m_N}{E_0} \, V_{1\pi}^{\rm static} \, \frac{m_N}{E_0}  = 
\left(1 -  \frac{p \, ^2 + p \,' {}^2}{2 m_N^2}  + \mathcal{O}(m_N^{-4})\right)    V_{1 \pi}^{\rm static}\,,
\eeq 
where $E_0 = \sqrt{p\, ^2 + m_N^2}$ is an operator.  Notice that here
and in what follows, we use the notation $p \equiv | \vec p \, |$, $p'
\equiv | \vec p \,' |$  
and $q \equiv | \vec q \, |$. The corresponding $1/m_N$-corrections to the TPEP 
read \cite{Friar:1999sj}\footnote{Notice that there are misprints in Eq.~(2.23) of Ref.~\cite{Epelbaum:2004fk} for $V_T^{(4)}$ and  $W_{T,S}^{(4)}$.}
\beqa
\label{tpe1m}
V_C^{(4)}&=& \frac{3 g_A^4}{512  \pi m_N F_\pi^4} \bigg\{ \frac{2
  M_\pi^5}{4 M_\pi^2 + q^2}-
 3  ( 4 M_\pi^4 - q^4 )  A (q) \bigg\}\,,\nonumber \\
W_C^{(4)} &=& \frac{g_A^2}{128 \pi m_N F_\pi^4} \Bigg\{ \frac{3 g_A^2
  M_\pi^5}{4 M_\pi^2 + q^2} 
- \bigg[ 4 M_\pi^2 + 2 q^2 - g_A^2 \left( 7 M_\pi^2 + \frac{9}{2} q^2 \right) \bigg] (2 M_\pi^2 + q^2 )
  A (q) \Bigg\}\,, \nn
V_T^{(4)} &=& -\frac{1}{q^2} V_S^{(4)}  =  \frac{9 g_A^4}{512 \pi m_N F_\pi^4}  \left(
4 M_\pi^2 + \frac{3}{2} q^2 \right) A (q) \,, \nn
W_T^{(4)} &=& -\frac{1}{q^2} W_S^{(4)} = - \frac{g_A^2}{256 \pi m_N F_\pi^4}  \left[
8 M_\pi^2 + 2 q^2 - g_A^2 \left( 4 M_\pi^2 + \frac{3}{2} q^2 \right) \right]  A (q) \,, \nn
V_{LS}^{(4)} &=& - \frac{3 g_A^4}{64 \pi m_N F_\pi^4} (2 M_\pi^2 + q^2 ) A (q) \,, \nn
W_{LS}^{(4)} &=& - \frac{g_A^2 ( 1 - g_A^2)}{64 \pi m_N F_\pi^4} 
(4 M_\pi^2 + q^2 ) A (q) \,.
\eeqa

It is customary to rewrite the relativistic Schr\"odinger equation
(\ref{schroed_rel}) in the equivalent non-relativistic form  \cite{Friar:1999sj} 
\beq
\label{schroed_nonrel}
\left[ \frac{p^2}{m_N}+ \tilde V \right] \Psi '
=  \frac{k \, ^2}{m_N} \Psi ' \,,
\eeq
where the potential operator $\tilde V$ is given by 
\beq
\label{rel_induced}
\tilde V = \bigg\{ \frac{\sqrt{p^2 + m_N^2}}{2 m_N}, \; V \bigg\} +
\frac{V^2}{4 m_N}\,,
\eeq 
and $\{ , \}$ denotes the anti-commutator.
This implies, in particular, that the $1/m_N$-corrections to the TPEP receive further
contributions induced by the second term in the right-hand side of
the above equation, $V_{1\pi}^2/(4 m_N)$, which have the form 
\beqa
\label{tpe1m_induced}
\delta V_C^{(4)}&=& \frac{3 g_A^4}{512  \pi m_N F_\pi^4}
( 2 M_\pi^2 + q^2 )^2  A (q)\,,\nonumber \\
\delta W_C^{(4)} &=& - \frac{g_A^4}{256  \pi m_N F_\pi^4}
( 2 M_\pi^2 + q^2 )^2  A (q)\,,\nonumber \\
\delta V_T^{(4)} &=& -\frac{1}{q^2} \delta V_S^{(4)}  =  - \frac{3 g_A^4}{1024  \pi m_N F_\pi^4}
( 4 M_\pi^2 + q^2 )  A (q)\,,\nonumber \\
\delta W_T^{(4)} &=& -\frac{1}{q^2} \delta W_S^{(4)} = \frac{g_A^4}{512  \pi m_N F_\pi^4}
( 4 M_\pi^2 + q^2 )  A (q)\,,
\eeqa
and need to be added to the expressions in Eq.~(\ref{tpe1m}).  It is
this form of the Schr\"odinger equation which was used in the NPWA and
is employed in the present analysis. We refer the reader to  appendix \ref{sec:scatt} for more
details on the kinematics and notations. To summarize,  the
relativistic corrections to the NN potential at N$^3$LO in the 
cms employed in the present work consist of: 
\begin{itemize}
\item
$1/m_N$-corrections to the static TPEP according to Eqs.~(\ref{tpe1m}) and
(\ref{tpe1m_induced}),
\item
$\mathcal{O}(m_N^{-2})$-corrections to the static OPEP and the
resulting TPEP according to Eqs.~(\ref{1pep_rel}) and
(\ref{rel_induced}), 
\beq
\bigg\{ \frac{m_N}{2E_0}, \;  V_{1\pi}^{\rm static} + V_{2\pi} \bigg\}
- ( V_{1\pi}^{\rm static} + V_{2\pi} ) \,.
\eeq 
This particular choice is appropriate at the order we are working and
is well suited for the local regularization of the long-range
potentials employed in our analysis,
see the next section for more details. Notice that the relativistic
corrections to the contact interactions in the cms have the same
form as the static terms and thus need not to be considered
separately.  
\end{itemize}
We emphasize that we also included the leading $1/m_N$-corrections
emerging from triangle two-pion exchange diagrams involving subleading
$\pi N$ vertices, i.e.~proportional to $c_i$.  The corresponding
expressions are not affected by unitary ambiguity of the potential and
can be found in Ref.~\cite{Kaiser:2001pc}. While these
contributions appear nominally at
next-to-next-to-next-to-next-to-leading order in the chiral expansion
within the employed power counting scheme, the resulting potentials
are known to be rather strong, presumably due to the LECs $c_{2,3,4}$
being numerically large. We have checked that neglecting those
terms does not substantially affect the quality of the fits but would
result in a smaller range of cutoffs.  

\section{Regularization}
\def\theequation{\arabic{section}.\arabic{equation}}
\label{sec:Regular}

Nuclear potentials derived in chiral EFT generate ultraviolet (UV)
divergences once substituted into the Lippmann-Schwinger (LS)
equation. The appearance of UV divergences in loop diagrams is an
intrinsic feature of any EFT which can be traced back to the derivative
expansion of the effective Lagrangian. While perturbative
calculations of e.g.~pion-pion or pion-nucleon scattering  within
chiral perturbation theory are usually organized in such a way that
\emph{all} UV divergences at a given order are absorbable into 
redefinition of the available LECs, the situation is
different for nucleon-nucleon scattering described in terms of
non-perturbative solution of the LS equation. While it is possible to
formulate a \emph{renormalizable} approach to NN scattering with
non-perturbative treatment of the OPEP \cite{Epelbaum:2012ua,Epelbaum:2013ij,Epelbaum:2013naa}, a much
simpler and 
commonly adopted  way to renormalize the LS equation is based on 
introducing a finite UV cutoff. (Implicit) renormalization of the NN
amplitude is then achieved by tuning the (bare) LECs accompanying the
contact interactions to experimental data or phase shifts.
 One advantage of such an approach, beyond its simplicity, is
the ability to combine the resulting nuclear potentials with the
available few- and many-body machinery which allows one to access  
observables beyond the NN system. The obvious disadvantage compared to the
renormalizable framework suggested in Ref.~\cite{Epelbaum:2012ua} is the 
appearance of finite-cutoff artefacts as manifested e.g.~in 
a residual cutoff dependence of nuclear observables. This feature is
unavoidable in calculations within such an approach (unless one is able
to subtract \emph{all} divergent integrals generated by iterations of the
chiral potentials in the LS equation). As a consequence, the UV
momentum-space cutoff $\Lambda$ has to be kept finite and (ideally) of
the order of the pertinent breakdown scale in the problem 
\cite{Lepage:1997cs,Epelbaum:2006pt,Epelbaum:2009sd,Marji:2013uia}.  

In practice, one is rather
limited with respect to the range of 
cutoff values since choosing $\Lambda \sim M_\rho$ or larger was
already found
to result in spurious deeply-bound states
\cite{Epelbaum:2004fk}. While such
unphysical deeply bound state do not affect low-energy observables,
they do 
drastically complicate applications to three- and more-nucleon
systems. For this reason, Ref.~\cite{Epelbaum:2004fk} has employed the cutoff range of $\Lambda
= 450 \ldots 600\,$MeV while the Idaho N$^3$LO potential is available
for two cutoff values only, namely $\Lambda = 500\,$MeV and $\Lambda =
600\,$MeV 
\cite{Entem:2003ft,Machleidt:2011zz}.  We further emphasize that  lattice
spacings employed in recent nuclear lattice
simulations of
Ref.~\cite{Borasoy:2006qn,Epelbaum:2009pd,Epelbaum:2011md,Epelbaum:2012qn,Lahde:2013uqa,Epelbaum:2013paa}
correspond to even  smaller cutoff
values. 

Given the relatively low values of $\Lambda$,  it is clearly desirable, in order to increase the accuracy and
applicability range of nuclear chiral EFT, to reduce
the amount of finite-cutoff artefacts, see Ref.~\cite{Lu:2014xfa} for a recent
lattice EFT work in a similar spirit, or at least to employ
regularization which avoids introducing unnecessary
artefacts. In the following, we will
argue that the momentum-space regularization used in the N$^3$LO
potentials of Refs.~\cite{Epelbaum:2004fk,Entem:2003ft} does induce
certain kinds of artefacts which can be easily avoided by carrying out
regularization in coordinate space as used recently in the
construction of the local
chiral NN potentials up to next-to-next-to-leading order (N$^2$LO) \cite{Gezerlis:2013ipa,Gezerlis:2014zia}.

\begin{figure}[tb]
\vskip 1 true cm
\includegraphics[width=0.5\textwidth,keepaspectratio,angle=0,clip]{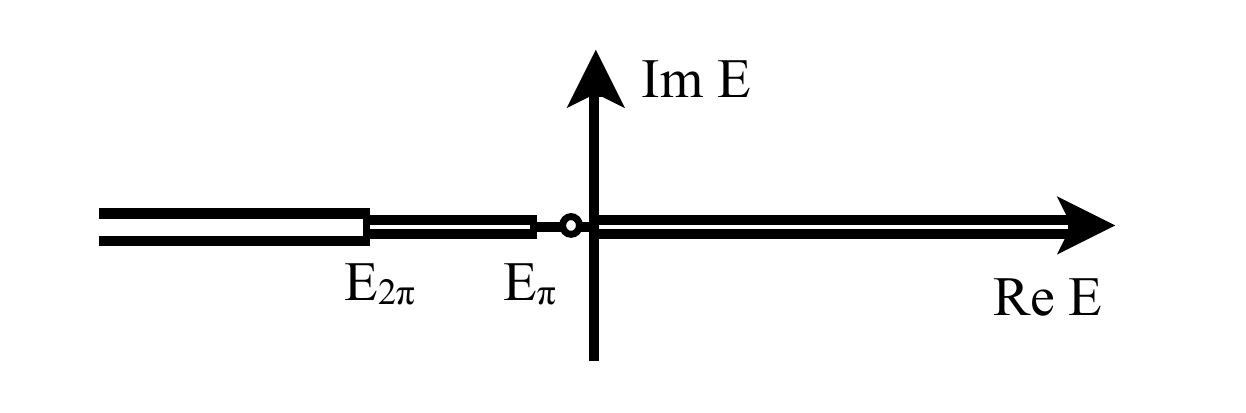}
    \caption{Singularity structure of the partial-wave two-nucleon
      scattering amplitude in the complex energy plane.  The solid dot
      indicates the position of the S-wave (virtual) bound
      state. Elastic unitarity is satisfied on the right-hand cut,
      also called unitarity cut. Left-hand cuts are caused by exchange
      processes in the potential. The first and second left-hand cuts
      due to one- and two-pion exchange start at laboratory energy of
      $E_\pi = - M_\pi^2/(2 m_N) \sim 10\,$MeV and  $E_{2\pi} = - 2
      M_\pi^2/m_N \sim 40\,$MeV, respectively.
\label{fig:cuts} 
 }
\end{figure}

Chiral nuclear forces involve generally two distinct kinds of contributions: first, at large
distances the potential is governed by contributions emerging from pion
exchanges which are
unambiguously\footnote{Strictly speaking, even the long-range tail of the potential is
  scheme-dependent as it can be affected by unitary
  transformations. Notice, however, that unitary ambiguity of the
  chiral nuclear forces was found to be strongly reduced in the static
  limit if one
  demands that the corresponding potentials are renormalizable \cite{Epelbaum:2007us}.} determined
by the chiral symmetry of QCD and experimental information on the
pion-nucleon system needed to pin down the relevant LECs. Secondly,
the short-range part of the potential is 
parametrized by all possible contact interactions with increasing
number of derivatives. It is desirable to introduce regularization in
such a way that the long-range part of the interaction including 
especially the OPEP, which is responsible for left-hand cuts in the
partial-wave scattering amplitude as visualized in Fig.~\ref{fig:cuts} and thus governs near-threshold energy
behavior of the S-matrix, is not affected by the regulator. Notice that
the near-threshold left-hand singularities of the amplitude can be
tested e.g.~via the low-energy theorems \cite{Cohen:1998jr,Epelbaum:2009sd}.  

The
standard implementation of the regulator used e.g.~in
Refs.~\cite{Epelbaum:2004fk,Entem:2003ft} is as follows:
\beq
\label{reg_stand}
V(\vec p \,', \vec p \, ) \to V_{\rm reg }(\vec p \,', \vec p \, ) =
V(\vec p \,', \vec p \, ) \; \exp \left(-\frac{{p'} ^m + p^m}{\Lambda^m}
  \right)\,,
\eeq
where the power $m$ is chosen sufficiently large in order that  the cutoff artefacts
$V(\vec p \,', \vec p \, )  \times \mathcal{O}\big( (Q/\Lambda)^m
\big)$ are beyond the chiral order one is working at. Specifically,
Ref.~\cite{Epelbaum:2004fk} used $m=6$ while Ref.~\cite{Entem:2003ft}
employed different powers $m\leq 6$ for different terms in the potential,
presumably in order to optimize the quality of the fit. It is clear
that the multiplicative regulator introduced above leads to
distortions of the analytic structure of the partial-wave amplitude near
threshold as it affects the discontinuity across the left-hand
cuts, see also Refs.~\cite{Gasparyan:2012km,Oller:2014uxa} for recent
studies of NN scattering
which explicitly exploit the analytic structure of the amplitude.
 While such distortions are small if $\Lambda$ can be chosen
sufficiently large, they can lead to sizable effects for the commonly
adopted choices of $\Lambda \sim 500\,$MeV. 
It is easy to avoid this unpleasant feature by exploiting the fact
that long-range potentials derived in chiral EFT are nearly
local, i.e.~depend only on momentum transfer $\vec q$. In fact, the only source
of non-locality is given by relativistic corrections which, in the
power counting scheme we are using, start to appear at N$^3$LO, see the
previous section. The feature of locality naturally suggests to apply 
regularization in coordinate space similar to what was done in
Refs.~\cite{Gezerlis:2013ipa,Gezerlis:2014zia} by cutting off
\emph{short-range} parts of the pion-exchange potentials, for which
chiral expansion does not converge, see Ref.~\cite{Baru:2012iv} for a related
discussion:
\beq
V_{\rm long-range} ( \vec r \, )  \to V_{\rm long-range}^{\rm reg} (
\vec r \, ) = V_{\rm long-range} ( \vec r \, ) f \left( \frac{r}{R}
\right)\,,
\eeq
where the regulator function $f(x)$ is chosen such that its value  goes to $0$ $(1)$ sufficiently fast
for $x \to 0$ (exponentially fast for $x \gg 1$).  It is instructive
to write this regularization in momentum space,
\beq
V ( \vec q \, )  \to V^{\rm reg} ( \vec q \, ) = V ( \vec q \, )  -
\int \frac{d^3 l}{(2 \pi)^3} V (\vec l \, ) \, {\rm FT}_{\vec q - \vec l}  \left[ 1 - f
\right] \,,
\eeq
where ${\rm FT}$ stays for the Fourier-Transform.  Given that ${\rm
  FT} \left[ 1 - f \right]$ is a short-range operator, the second term in
the right-hand side of the above equation does not induce any 
long-range contributions. This is in contrast to the 
procedure specified in Eq.~(\ref{reg_stand}), where long-range terms
are induced by regularization (albeit suppressed by inverse powers
of the cutoff $\Lambda$). Notice further that the suggested regularization is
qualitatively similar to the well-known Pauli-Villars
regularization. 

As will be shown in the next section, the above choice of the
regulator
makes the additional SFR of the pion
exchange contributions obsolete. This is a particularly welcome feature
in view of the fact that the expressions for the three-nucleon force
at N$^3$LO \cite{Ishikawa:2007zz,Bernard:2007sp,Bernard:2011zr}
are only available in the framework of DR
which corresponds to choosing an infinitely large cutoff in the spectral function
representation. Notice further that recent calculations of the  three-nucleon force
beyond N$^3$LO \cite{Krebs:2012yv,Krebs:2013kha} are also carried out in
the framework of DR.  
The SFR was originally introduced in
Refs.~\cite{Epelbaum:2003gr,Epelbaum:2003xx} as an attempt to avoid
unnaturally strong attraction generated by the subleading TPEP 
in the isoscalar central channel \cite{Kaiser:1997mw} caused by short-range 
components in the spectral representation.   
The SFR framework was used to construct the N$^3$LO potential of
Ref.~\cite{Epelbaum:2004fk}. Notice, however,  that in spite of employing
the SFR, it was necessary to set the LEC $c_3$, which governs the isoscalar part of the N$^2$LO
TPEP,  to its lowest  in magnitude
value still compatible with $\pi N$ scattering  in
order to avoid the  appearance of deeply bound  states. 

We are now in the position to specify the regulator function
$f(r/R)$, for which the choice $f (r/R) = 1- \exp (-(r/R)^4)$ was
adopted in Refs.~\cite{Gezerlis:2013ipa,Gezerlis:2014zia}. Given that
DR expressions for TPEP at N$^2$LO behave at short distances as $1/r^6$, such a regulator is
insufficient to make the DR potential non-singular and can only be used
in combination with the SFR which makes the TPEP less singular. Notice
further that such a 
regulator induces oscillations in momentum-space matrix elements of
the potential $V(\vec q \,)$  for large values of $q$ which may represent
a considerable complication for numerical applications. In order to
avoid this unpleasant feature, the regulator function can be chosen in
the form 
\beq
\label{NewReg}
f \left( \frac{r}{R} \right) = \left[1- \exp\left( - \frac{r^2}{R^2} \right)\right]^n\,,
\eeq
where the exponent $n$ has to be taken sufficiently large. It is
necessary to choose $n = 4$ or larger in order to make the regularized
expressions for the DR TPEP at N$^3$LO vanishing in the
origin, but we found that larger values of $n$ lead to more stable
numerical results when doing calculations in momentum
space.\footnote{Given that locally-regularized potentials $V(\vec q
  \,)$ show only a 
power-low decrease for high values of momentum transfer $q$, much
higher virtual momenta are involved in solving the LS equation as compared to
the nonlocal chiral potentials of Refs.~\cite{Epelbaum:2004fk,Entem:2003ft}.} 
Here and in what follows, we make the choice $n=6$. 
For contact interactions, we employ the standard nonlocal regulator
specified  in Eq.~(\ref{reg_stand}) and set $m=2$ so that the
regulator is again of a Gaussian type. In order to have a single
cutoff scale, we will relate the coordinate- and momentum-space
cutoffs $R$ and $\Lambda$ by setting $\Lambda = 2R^{-1}$ motivated by
the relation ${\rm FT}_q \big[\exp (-r^2/R^2) \big] \propto \exp (-q^2
R^2/4)$. We will show below that the results of our analysis depend
little on specific details of the regulator.

\section{Fits and results for the phase shifts}
\def\theequation{\arabic{section}.\arabic{equation}}
\label{sec:PhaseShifts}

Having specified the regularization, we now describe the fit
procedure and show our results for phase shifts.  We begin with
specifying the values of the LECs and masses that enter the
potentials. Here and in what follows, we use $m_p=938.272\,$MeV,
$m_n=939.565\,$MeV for the nucleon masses and $M_{\pi^\pm}= 139.57\,$MeV
and $M_{\pi^0}= 134.98\,$MeV for the charged and neutral pion masses,
respectively. For the average pion mass which enters the expressions
for the TPEP the value $M_{\pi}= 138.03\,$MeV is adopted. Further, we
use the values $F_{\pi}=92.4\,$MeV and  $g_A=1.267$ for the pion decay
and nucleon axial coupling constants. Starting from NLO, one needs to
account for the  Goldberger-Treiman discrepancy which can be achieved
via the replacement 
\beq
g_A \to g_A - 2 d_{18} M_\pi^2\,,
\eeq
where $d_{18}$ is a LEC from the sub-subleading pion-nucleon
effective Lagrangian. Following \cite{Epelbaum:2004fk}, we adopt the larger value 
$g_A = 1.29$ instead of $g_A=1.267$ in order to account for the
Goldberger-Treiman discrepancy in the expressions for the OPEP and, at N$^3$LO, also for the leading
TPEP. Using the Goldberger-Treiman relation
$g_{\pi N} = g_A m_N/F_\pi$, this value of $g_A$ leads to $g_{\pi
  N}^2/(4 \pi) = 13.67$ which is consistent with  the recent
determination via the Goldberger-Miyazawa-Oehme sum rule \cite{Baru:2010xn},  $g_{\pi
  N}^2/(4 \pi) = 13.69 \pm 0.20$, as well as with the  older
determinations from NN \cite{deSwart:1997ep}  and $\pi N$ \cite{Arndt:1994bu} scattering data. 

It remains to specify the $\pi N$ LECs $c_i$ and $d_i$ which enter the
TPEP at N$^3$LO. In Table \ref{tab_c_i}, we list the values of the
$c_i$'s adopted in the N$^3$LO potentials of Refs.~\cite{Epelbaum:2004fk,Entem:2003ft} and in the
current work together with the empirical values determined from
pion-nucleon scattering inside the Mandelstam triangle (fit 1). 
\begin{table}[t]
\caption{Values of the LECs $c_i$ in units of GeV$^{-1}$ used in the
  various N$^3$LO NN potentials in comparison with the empirically
  determined values from $\pi N$ scattering as described in the text.  
\label{tab_c_i}}
\smallskip
\begin{tabular*}{\textwidth}{@{\extracolsep{\fill}}rrrrr}
\hline 
\hline 
\noalign{\smallskip}
LEC  &  N$^3$LO potential of Ref.~\cite{Entem:2003ft} &   N$^3$LO potential of Ref.~\cite{Epelbaum:2004fk}
&   this work
& Empirical  
\smallskip
 \\
\hline 
\hline 
$c_1$ & $-0.81 \; \,$ & $-0.81 \; \,$  & $-0.81$  & $-0.81\pm 0.15$ \cite{Buettiker:1999ap} \\ 
$c_2$ &   $2.80^\star$ &   $3.28 \; \,$  &   $3.28$  &   $3.28\pm
0.23$ \cite{Fettes:1998ud} \\ 
$c_3$ & $-3.20^\star$ &$-3.40^\dagger$  & $-4.69$  & $-4.69\pm 1.34$ \cite{Buettiker:1999ap}  \\ 
$c_4$ &  $5.40^\star$ &  $3.40 \; \,$  & $3.40$ & $3.40\pm 0.04$ \cite{Buettiker:1999ap} \\  [4pt]
%\smallskip
\hline 
\hline 
\multicolumn{5}{l}{$^\star$Fit parameter.} \\ 
\multicolumn{5}{l}{$^\dagger$Larger in magnitude values were found to
  lead to spurious deeply-bound states. } \\ 
\end{tabular*}
\end{table}
Using this unphysical kinematics in combination with dispersion relations has the advantage that the chiral
expansion converges faster than in the physical
region. Thus, one expects the determined LECs to have smaller theoretical uncertainties due to
truncation of the chiral expansion as compared to fits in the physical
region. For the LEC $c_2$, which could not be reliably determined in
\cite{Buettiker:1999ap}, we give the value from the order $Q^3$
heavy-baryon calculation of Ref.~\cite{Fettes:1998ud}.  We emphasize
that several more recent determinations of these LECs from $\pi N$
scattering up to order $Q^4$ in the heavy-baryon as well as manifestly
covariant formulations of chiral perturbation theory are available,
see e.g.~\cite{Krebs:2012yv,Fettes:2000xg,Alarcon:2012kn,Chen:2012nx,Wendt:2014lja}.
In addition, attempts were made to determine the LECs $c_{1,3,4}$ from
nucleon-nucleon scattering data based on the two-pion exchange
potential calculated at N$^2$LO
\cite{Rentmeester:1999vw,Entem:2002sf,Rentmeester:2003mf,Entem:2003cs,Ekstrom:2014dxa,Perez:2014bua}
and N$^3$LO
\cite{Entem:2003ft}. In particular, the values found in Refs.~\cite{Rentmeester:1999vw,Rentmeester:2003mf,Perez:2014bua} are
consistent, within the quoted uncertainties, with the results obtained in
pion-nucleon scattering.  Notice,
however, that none of these studies have addressed the question of the \emph{systematic}
theoretical uncertainties, in particular due to truncation of the  chiral expansion for the
potential at a given order. Accordingly, the interpretation of these
findings is not completely clear. Finally, for the LECs $d_i$ from the
order-$Q^3$ effective pion-nucleon Lagrangian which contribute to the
N$^3$LO  TPEP we adopt, following Refs.~\cite{Epelbaum:2004fk,Entem:2003ft}, the central values from fit 1
to $\pi N$ phase shift given in Ref.~\cite{Fettes:1998ud}, namely
\begin{equation}    
\bar d_1 + \bar d_2 = 3.06 \mbox{ GeV}^{-2}, \quad \quad 
\bar d_3 = -3.27 \mbox{ GeV}^{-2}, \quad \quad 
\bar d_5 = 0.45 \mbox{ GeV}^{-2}, \quad \quad 
\bar d_{14}- \bar d_{15}  = -5.65 \mbox{ GeV}^{-2} \,.
\end{equation}

For the regulator $R$, we employ the same range as used in the local
versions of the N$^2$LO  NN potential of Ref.~\cite{Gezerlis:2013ipa},
namely $R=0.8 \ldots 1.2\,$fm. Specifically, we will carry out fits for
five different values of $R$, namely $R=0.8\,$fm, $R=0.9\,$fm,
$R=1.0\,$fm, $R=1.1\,$fm and $R=1.2\,$fm.   Notice that the smallest value of the
cutoff $R$, $R=0.8\,$fm, coincides with
the estimated distance at which the chiral expansion of the NN
potential is expected to break down \cite{Baru:2012iv}.  
When transformed to momentum space using
the relation $\Lambda = 2 R^{-1}$ as motivated in the previous section, 
the employed cutoff range corresponds to the range of $\Lambda \simeq 500
\ldots 330\,$MeV.

Following the procedure of the NPWA \cite{Stoks:1993tb} which we use
as input for our calculations, we fit all isospin-$1$ channels to pp phase shifts, which are accurately determined from
the available scattering data, and generate np and
nn phase
shifts (with the exception of the $^1$S$_0$ partial wave) by using the appropriate kinematical relations, see 
Eqs.~(\ref{kin_pp}), ~(\ref{kin_nn}) and ~(\ref{kin_np}), 
taking into account the isospin-breaking
corrections to the OPEP due to the different pion masses, see
Eq.~(\ref{ope_iso}), and switching off the long-range electromagnetic
interactions\footnote{Notice that such a ``minimalistic'' treatment of
IB effects in the present analysis is dictated by using
the NPWA results rather than NN experimental data as input for our
fits. In particular, we do not take into account the known IB 
two-pion exchange contributions since they would induce shifts
$\delta_{\rm np} - \delta_{\rm pp}$ incompatible with the results of
Ref.~\cite{Stoks:1993tb}. A more complete treatment of IB corrections
using NN experimental data is left for a future work.}. More precisely, pp phase shifts of
Ref.~\cite{Stoks:1993tb}, which we use as input in our fits,
actually correspond to phase shifts of the electromagnetic plus nuclear
interaction with respect to electromagnetic wave functions,
i.e.~$\delta_{{\rm EM} + N}^{\rm EM}$ in the notation of Ref.~\cite{Stoks:1993tb}. The dominant contributions to
the long-range electromagnetic interaction are well known and include
the usual static Coulomb potential, the leading relativistic correction
to the Coulomb potential, the magnetic moment interaction and the
vacuum polarization potential, see e.g.~\cite{Stoks:1993tb,Epelbaum:2004fk} and references
therein for more details and explicit expressions. Notice that the
static Coulomb potential in combination with the leading relativistic
correction is often referred to as the modified or relativistic
Coulomb interaction. The NPWA employs the
approximation $\delta_{{\rm EM} + N}^{\rm EM} \approx \delta_{{\rm C}
  + N}^{\rm C}$ for all pp channels except $^1$S$_0$, where
$C$ means that the electromagnetic interaction is approximated by 
the
Coulomb potential. For the  $^1$S$_0$ partial wave, the approximate relation
between the phase shifts $\delta_{{\rm EM} + N}^{\rm EM}$ published in \cite{Stoks:1993tb}
and $\delta_{{\rm C} + N}^{\rm C}$ can be obtained using distorted
wave Born approximation. The resulting fairly model-independent shifts are tabulated in
Ref.~\cite{Bergervoet:1988zz}  and appear to be negligibly
small for energies larger than $E_{\rm lab} \sim 30\,$MeV. Throughout
this work, we use pp phase shifts corresponding to the modified
Coulomb plus nuclear interaction with respect to the modified Coulomb
wave functions and employ the corrections given in
Ref.~\cite{Bergervoet:1988zz} to relate these phase shifts to the
ones of the NPWA \cite{Stoks:1993tb} in the  $^1$S$_0$ partial wave. 
For the np case, the calculated and shown 
phase shifts are that of the nuclear interaction with respect to
Ricatti-Bessel functions. Notice that np phase shifts of
the NPWA \cite{Stoks:1993tb} actually correspond to $\delta_{{\rm MM}
  + N}^{\rm MM}$ since  the electromagnetic interaction in that case is entirely
given by the magnetic moment (MM) interaction within the NPWA. 
It is well known that the approximation \cite{Wiringa:1994wb} $\delta_{{\rm EM} + N}^{\rm EM}
\approx \delta_{N}$ is rather accurate for  all channels except for the $^3$S$_1$ partial
wave \cite{Wiringa:1994wb}. We will employ this standard approximation
for \emph{all} np partial waves in order to directly compare our
phase shifts with the ones of the NPWA \cite{Stoks:1993tb}. It should
be understood that effects of the magnetic moment interaction in $\delta_{{\rm MM} + N}^{\rm
  MM}$ in the $^3$S$_1$ channel of the NPWA  \cite{Stoks:1993tb} are mimicked by contact
interactions in the potential when we calculate $\delta_{N}$. We
emphasize that our choice for the phase shifts throughout this work is
the same as the one adopted e.g. in
Refs.~\cite{Entem:2003ft,Machleidt:2000ge}. 

For each value of the cutoff $R$, we determine the LECs accompanying
the short-range operators specified in Eq.~(\ref{VC}) from a fit to
np and pp phase shifts of the NPWA \cite{Stoks:1993tb}. 
Specifically, we have in total 4 LECs  at LO ($\tilde C_{3S1}$,  $\tilde C_{1S0}^{\rm
  np}$, $\tilde C_{1S0}^{\rm pp}$ and $\tilde C_{1S0}^{\rm nn}$),
11 LECs at NLO and N$^2$LO   ($\tilde C_{3S1}$,  $\tilde C_{1S0}^{\rm
  np}$, $\tilde C_{1S0}^{\rm pp}$, $\tilde C_{1S0}^{\rm nn}$ and $C_i$)
and 26 LECs at N$^3$LO ($\tilde C_{3S1}$,  $\tilde C_{1S0}^{\rm
  np}$, $\tilde C_{1S0}^{\rm pp}$, $\tilde C_{1S0}^{\rm nn}$, $C_i$ and
$D_i$). Notice that while it is not necessary to
account for isospin breaking at LO from the point of view of
power counting, we decided to include the same isospin-breaking
corrections in order to
be consistent with the procedure of the NPWA and to allow for a
meaningful comparison of results at different orders. 
The fits are
carried out using the same energies as employed in the multi-energy partial
wave analysis of the Nijmegen group, namely $E_{\rm lab} = 1$, $5$,
$10$, $25$, $50$,  $100$,  $150$, $200$, $250$ and $300\,$MeV.
Specifically, we use the energies  $E_{\rm lab} \leq 25\,$MeV at LO,   
$E_{\rm lab} \leq 100\,$MeV at NLO and N$^2$LO and  $E_{\rm lab} \leq
200\,$MeV at N$^3$LO. The results for phase shifts at higher energies
are thus to be regarded as predictions. 

At N$^3$LO, we found that
fits in the $^3$S$_1$-$^3$D$_1$ become unstable, which manifests itself
in the appearance of different solutions which describe equally well the $^3$S$_1$ and
$^3$D$_1$ phase shifts and the mixing angle $\epsilon_1$. This feature
becomes especially disturbing for the hardest cutoff choices of $R=0.8\,$fm and  $R=0.9\,$fm and indicates that 8 unknown LECs in
this channel offer too much flexibility in the description of the phase
shifts and the mixing angle. In addition to requiring that the resulting LECs are of
natural size, we decided to impose further constraints to stabilize
the fits in this channel. In particular, we demand that the deuteron
binding energy is correctly reproduced and discard solutions which
lead to unrealistic values of the $D$-state probability in the
deuteron or show a too strong violation of the Wigner SU(4) symmetry which
implies the relation $\tilde C_{1S0} \simeq \tilde C_{3S1}$
\cite{Mehen:1999qs}, see also  Ref.~\cite{Epelbaum:2001fm}.  
It should be emphasized that the deuteron $D$-state
  probability $P_D$ is not a measurable quantity and can be changed by
  means of a unitary transformation \cite{Friar:1979zz}. 
Modern phenomenological NN potentials typically yield the values of 
$P_D$ in the range of $P_D = 4 \ldots 6$\%. In particular, the 
AV18 \cite{Wiringa:1994wb}, Nijmegen I and II and Reid93 \cite{Stoks:1994wp} potentials have $P_D=5.76$\%, $P_D=5.66$\%
$P_D=5.64$\% and  $P_D=5.70$\%, respectively, while the CD-Bonn
potential \cite{Machleidt:2000ge} leads to a  smaller value of $P_D=4.85$\%, see Ref.~\cite{Forest:1999wu}
for a related discussion. Furthermore,  the chiral N$^3$LO potential
of Refs.~\cite{Entem:2003ft} yields $P_D=4.51$\% while the ones of
Ref.~\cite{Epelbaum:2004fk} lead to even smaller values. It
is conceivable  that NN potentials corresponding to the choice of unitary
transformation leading to values of $P_D$ very different from the ones 
listed above would require strong many-body forces and exchange
currents, the feature which is certainly worrisome in the context of
effective field theory but also from the computational point of view,
see e.g.~the discussion in Ref.~\cite{Machleidt:2009bh}. Thus, we
decided to introduce the deuteron D-state probability $P_D=5\% \pm 1\%$
as an additional ``data'' point in the fit.  
%This is a rather weak constraint given that the 
%total $\chi^2$ for reproducing  the $^3$S$_1$ and   $^3$D$_1$ phase
%shifts and the mixing angle $\epsilon_1$ at eight energies as described
%above is in the range of $3.3 \ldots 7.6$ (with largest value
%corresponding to the softest choice of the cutoff).\footnote{Our definition of
 % the uncertainties entering the calculation of $\chi^2$ will be given
%below.} Nevertheless, imposing this constraint on the value of $P_D$
%leads to a considerable stabilization of the fit results.   
As for the
second constraint on the value of $\tilde C_{3S1}$, we employ a
simple ``augmented $\chi^2$" following the lines of
Ref.~\cite{Schindler:2008fh} to penalize those values of $\tilde
C_{3S1}$ which are considerably different from the ones of $\tilde C_{1S0}$
for the same choice of the cutoff. In practice, this is achieved by
using 
\begin{equation}
\chi_{\rm aug}^2 = \chi^2 + \chi_{\rm prior}^2\,, \quad \quad
\mbox{with} \quad   \chi_{\rm prior}^2 = \frac{(\tilde C_{3S1} - \tilde
C_{1S0} )^2}{(\Delta \tilde C_{3S1})^2}\,,
\end{equation}
where we choose $\Delta \tilde C_{3S1} =  \tilde C_{1S0} /4$.  Notice
that this additional constraint is, in fact, only active for the two
hardest choices of the cutoff as the unconstrained fits in other cases
already lead to $\tilde C_{3S1} \simeq \tilde C_{1S0}$. For example,
the value for $\tilde C_{3S1}$ resulting from the unconstrained fit
with the cutoff $R=1\,$fm appears to lie within $1 \%$ of the value of  $\tilde
C_{1S0}$. 

Having specified the details of the fitting procedure, we are now in the position
to discuss the results. In Table \ref{tab_LEC}, we give the obtained values
of the various LECs at N$^3$LO for different choices of the cutoff $R$. 
\begin{table}[t]
\caption{The LECs $\tilde C_{1S0}^{\rm pp}$,   $\tilde C_{1S0}^{\rm
    nn}$,  $\tilde C_{1S0}^{\rm np}$,  $\tilde C_{3S1}$, $C_i$ and
  $D_i$ at N$^3$LO for different values of the cutoff $R$. The values
  of the $\tilde C_i$, $C_i$ and $D_i$ are given in units of $10^4$
  GeV$^{-2}$, $10^4$
  GeV$^{-4}$ and $10^4$
  GeV$^{-6}$, respectively.
\label{tab_LEC}}
\smallskip
\begin{tabular*}{\textwidth}{@{\extracolsep{\fill}}rrrrrr}
\hline 
\hline
\noalign{\smallskip}
LEC &  $R=0.8\,$fm &  $R=0.9\,$fm &  $R=1.0\,$fm &  $R=1.1\,$fm &  $R=1.2\,$fm 
\smallskip
 \\
\hline 
\hline
%\smallskip
$\tilde C_{1S0}^{\rm pp}$   &  $0.2363$   & $0.1648$       & $0.1090$    & $0.0574$  & $0.0166$  \\
$\tilde C_{1S0}^{\rm nn}$   &  $0.2352$   & $0.1631$       & $0.1070$    & $0.0551$  &  $0.0140$ \\
$\tilde C_{1S0}^{\rm np}$   &  $0.2328$   & $0.1600$       & $0.1035$    & $0.0513$  &  $0.0100$ \\
$C_{1S0}$                         &  $-0.0433$ & $-0.0400$   & $-0.1155$  & $-0.2078$  &  $-0.3344$ \\
$D_{1S0}^1$                     &  $3.0691$   & $-1.8080$  & $-6.9812$  & $-13.3659$  & $-21.9934$  \\
$D_{1S0}^2$                     &  $0.6135$   & $3.0136$      &     $6.4537$& $10.8060$  & $16.5111$  \\
$C_{3P0}$                         &  $1.0678$   & $0.7812$      &    $0.5120$  & $0.2645$  &  $0.0313$ \\
$D_{3P0}$                         &  $-0.4030$ & $0.6311$      &    $1.5809$  & $2.5084$  &  $3.4432$ \\
$C_{1P1}$                         &  $1.0068$   & $0.8095$      &    $0.6926$  & $0.6267$  &  $0.5930$ \\
$D_{1P1}$                         &  $0.9480$   & $1.6652$      &    $2.4623$  & $3.4613$  &  $4.7819$ \\
$C_{3P1}$                         &  $1.3413$   & $1.1253$      &    $0.9667$  & $0.8592$  &  $0.7865$ \\
$D_{3P1}$                         &  $-0.7070$ & $0.3656$      &    $1.4145$  & $2.6058$  &  $4.1340$ \\
$\tilde C_{3S1}$               &  $0.2441$   &  $0.1688$      &    $0.1043$  & $0.0513$  &  $0.0097$ \\
$C_{3S1}$                         &  $-0.3292$ & $-0.3844$  &  $-0.4256$  & $-0.4868$  &   $-0.5790$\\
$D_{3S1}^1$                     &  $-4.5205$ & $-8.0894$ & $-12.3514$  & $-17.5859$  & $-23.6921$  \\
$D_{3S1}^2$                     &  $3.8438$   & $6.4034$    &     $9.2050$  & $12.3977$  &  $15.7636$ \\
$C_{3D1-3S1}$                   &  $0.3424$   & $0.4092$      &    $0.5388$  & $0.7298$  &  $0.9624$ \\
$D_{3D1-3S1}^1$               &  $0.8641$   & $-0.3181$  &  $-2.0898$  & $-4.3925$  &  $-7.3943$ \\
$D_{3D1-3S1}^2$               &  $-1.5054$ & $-0.3157$    &    $1.6466$  & $4.2798$  &  $8.0089$ \\
$D_{3D1}$                         &  $1.4422$  & $1.2225$      &    $1.1240$  & $1.1446$  &  $1.1740$ \\
$D_{1D2}$                         &  $1.3770$  & $0.9617$    &    $0.4782$  & $-0.1144$  &  $-0.8569$ \\
$D_{3D2}$                         &  $0.6540$  & $0.0259$    &  $-0.8805$  & $-2.1386$  &  $-3.8116$ \\
$C_{3P2}$                         &  $0.5639$   & $0.3189$      &    $0.1418$  & $0.0134$  &  $-0.0768$ \\
$D_{3P2}$                         &  $-0.5008$ & $-0.4398$   & $-0.6095$  & $-1.0773$  &  $-1.9421$ \\
$D_{3F2-3P2}$                   &  $-0.1355$ & $-0.2343$   & $-0.4108$  &  $-0.6946$ &  $-1.1275$ \\
$D_{3D3}$                         & $-0.1655$  & $-0.4103$   & $-0.7289$  & $-1.1377$  &  $-1.6564$  \\[4pt]
\hline \hline
\end{tabular*}
\end{table}
It is important to keep in mind that the LECs correspond to bare
quantities and are expected to depend significantly on the chiral order, employed
regularization scheme and the choice of the cutoff. 
Although the LECs at NLO and N$^2$LO were demonstrated
in Ref.~\cite{Epelbaum:2001fm} to be well described in terms of resonance saturation
by heavy-meson exchanges, it makes generally little sense to directly compare
the LECs obtained by using different regularization schemes with each
other. For example, due to the choice of a Gaussian regulator for
contact interactions adopted in the present work, the LECs $D_i$ contain  
contributions induced by contact interactions at lower orders driven
by the LECs $\tilde C_i$ and $C_i$. A more meaningful comparison between
the different approaches should
rather be done at the level of observables or, more generally,
renormalized quantities. What is, however, important to verify is that
the obtained LECs are of a natural size. The natural size for the
various LECs can be roughly estimated as \cite{Epelbaum:2004fk}
\begin{equation}
| \tilde C_i | \sim \frac{4 \pi}{F_\pi^2}, \quad \quad
| C_i | \sim \frac{4 \pi}{F_\pi^2 \Lambda_b^2}, \quad \quad
| D_i | \sim \frac{4 \pi}{F_\pi^2 \Lambda_b^4},
\end{equation}
where the factor of $4\pi$ emerges from the angular integration in the
partial wave decomposition and  $\Lambda_b$ is the pertinent hard
scale. If the scale  $\Lambda_b$ is identified with the employed ultraviolet cutoff 
$\Lambda = 2R^{-1}$,  the expected natural size of  the LECs $| \tilde C_i |$, $| C_i |$ and  $|
D_i |$ is $0.15 \times 10^4\,$GeV$^{-2}$, $0.6 \times
10^4\,$GeV$^{-4}$ ($1.4 \times 10^4\,$GeV$^{-4}$)
and $2.5 \times 10^4\,$GeV$^{-6}$ ($13 \times
10^4\,$GeV$^{-6}$), respectively, for the hardest (softest) employed cutoff $R=0.8\,$fm
($R=1.2\,$fm). This would imply that all obtained LECs are of a natural
size. On the other hand, as we will show in the next section,  the actual
breakdown scale $\Lambda_b$ in our case appears to be somewhat larger
than the UV cutoff $\Lambda = 2R^{-1}$. In particular, we will use 
$\Lambda_b = 400\ldots 600\,$MeV, depending on the employed value of $R$,
for estimating the theoretical uncertainty in section
\ref{sec:Uncertainty}. This implies that the natural size of the LECs $| C_i |$ and  $|
D_i |$ is expected to be $0.4 \times
10^4\,$GeV$^{-4}$ ($0.9 \times 10^4\,$GeV$^{-4}$)
and $1.1 \times 10^4\,$GeV$^{-6}$ ($6 \times
10^4\,$GeV$^{-6}$), respectively, for the hardest (softest) cutoff
choices. Also for such an estimation, all LECs are of a natural size
(with the values of $D_{3S1}^1$ appearing to be somewhat large in magnitude).  

\begin{figure}[tb]
\vskip 1 true cm
\includegraphics[width=\textwidth,keepaspectratio,angle=0,clip]{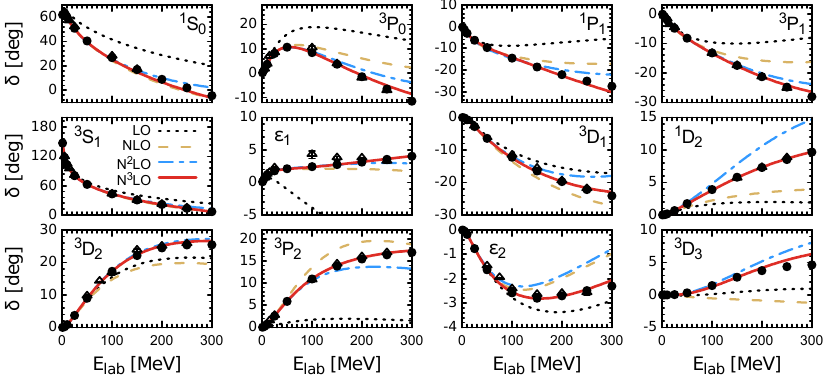}
    \caption{Chiral expansion of the NN phase shifts in comparison with the NPWA \cite{Stoks:1993tb}  (solid dots) and the GWU 
      single-energy np partial wave analysis \cite{Arndt:1994br} (open
      triangles).  Dotted, dashed (color online: brown), dashed-dotted
      (color online: blue) and solid (color online: red) lines show the
      results at LO, NLO, N$^2$LO and N$^3$LO, respectively, calculated using the cutoff $R=0.9\,$fm.  Only those partial waves are shown which 
     have been used in the fits at N$^3$LO. 
\label{fig:phases_conv} 
 }
\end{figure}

In Fig.~\ref{fig:phases_conv}, we show our results at different orders
in the chiral expansion for np
phase shifts and mixing angles used in the N$^3$LO fit.   
Here, we restrict ourselves to one particular cutoff choice, namely
$R=0.9\,$fm, in order to
have not too many lines in the plots. The cutoff dependence of our results
will be addressed in sections \ref{sec:CutDep} and \ref{sec:Uncertainty}.  One clearly observes a
good convergence pattern with the N$^3$LO results being in
excellent agreement with the NPWA in the whole considered range of energies. 
The convergence pattern is in most cases qualitatively similar to the one
reported in Ref.~\cite{Epelbaum:2004fk} although there are differences
in certain channels. For example, for the $^3$P$_0$ partial wave, the results at NLO
and N$^2$LO of Ref.~\cite{Epelbaum:2004fk} indicate too much repulsion
at higher energies while the opposite is observed in our 
analysis. Still, these results are consistent with each other within the estimated
theoretical uncertainty at these orders, see
section \ref{sec:Uncertainty} for more details. Concerning the  N$^3$LO results,
the improved NN potential of this work shows a superior 
performance in the whole considered energy range compared to the potential of
Ref.~\cite{Epelbaum:2004fk} as will be shown below. We attribute this feature primarily to a
better choice of regularization, see  section  \ref{sec:Regular} for
more details. We also emphasize that our N$^3$LO results for
peripheral, i.e.~F- and higher partial waves not shown in
Fig.~\ref{fig:phases_conv} are similar to the ones reported in
Refs.~\cite{Epelbaum:2004fk,Entem:2014msa}. In particular, we also observe large
relative deviations for F-waves at higher energies. For example, for
the cutoff $R=0.9\,$fm, we find $\delta_{3F2}^{\rm np} = 3.1^\circ$  at
$E_{\rm lab} = 250\,$MeV to be compared with the NPWA result
$\delta_{3F2}^{\rm np} = 1.4^\circ$. Notice, however, that 
\emph{absolute} deviations from the NPWA for F-waves appear to be of a
similar size as the ones observed in low partial wave so that there is
no reason to expect the theoretical uncertainty in low-energy observables to be
dominated by the discrepancies in F-waves. It is conceivable that
the deviations for F-waves will be largely reduced by the order-$Q^6$
contact interactions. 

It is desirable to have a quantitative criterium for comparing the
accuracy of different potentials with each other. Usually, this is
achieved by calculating the $\chi^2/{\rm datum}$ for the reproduction
of the available np and pp scattering data. 
Presently, we do not have the necessary machinery to carry out such a
calculation and reserve this task for a future study. In the present
work, we employ a simpler approach and calculate $\chi^2/{\rm datum}$
for the reproduction of the \emph{phase shifts} of the NPWA used as
input in our analysis.  Specifically, we calculate  $\chi^2/{\rm datum}$
at energies of  $E_{\rm lab} = 1$, $5$,
$10$, $25$, $50$,  $100$,  $150$, $200$, $250$ and $300\,$MeV employed
in the NPWA and also in our fits. Unfortunately, the NPWA
\cite{Stoks:1993tb} only provide statistical errors which do not
include systematic uncertainties. In order to have a meaningful
definition of $\chi^2$, we define the uncertainty for a given phase
shift (or mixing angle) $\delta$ in the channel $X$ at a given energy as 
\begin{equation}
\Delta_X = \max \left( \Delta^{\rm NPWA}_X, \;\; | \delta_X^{\rm NijmI} -
  \delta_X^{\rm NPWA} |, \;\; | \delta_X^{\rm NijmII} -
  \delta_X^{\rm NPWA} |, \;\; | \delta_X^{\rm Reid93} -
  \delta_X^{\rm NPWA} | \right)\,,
\end{equation}
where  $ \delta_X^{\rm NPWA}$ and $\Delta^{\rm NPWA}_X$ refer to the 
phase shift (or mixing angle) in the channel $X$ and the corresponding
statistical error of the NPWA, respectively, while $\delta_X^{\rm
  NijmI}$,  $\delta_X^{\rm NijmI}$ and  $\delta_X^{\rm Reid93}$ denote
the results based on the Nijmegen I, II and Reid93 NN potentials of
Ref.~\cite{Stoks:1994wp}. These phenomenological potentials are constructed
using the same database as employed in the NPWA and have a nearly
optimal $\chi^2/{\rm datum}$ of  $1.03$. For this
reason, they have, in fact, been suggested as alternative partial wave analyses \cite{Stoks:1994wp}. 
While the above definition of uncertainties provides a reasonable
estimation, one should not overinterpret the resulting values
for $\chi^2/{\rm datum}$ calculated based on the NPWA phase shifts in the way
specified above.\footnote{There is no obvious relation
  between $\chi^2/{\rm datum}$  for the description of NPWA phase
  shifts and that for the description of real data. In particular, our
  simplistic approach does
  not take into account the fact that peripheral partial waves are
  less important for the description of low-energy observables.} 
In particular, there is no statistical interpretation of the value of
$\chi^2/{\rm datum}$. 
We, nevertheless, still find this approach useful for the sake of a
simple comparative analysis of the  accuracy of
different NN potentials.  We also used the errors defined above in all
our fits.  

In Table \ref{tab_chi2_1} we show the $\chi^2/{\rm datum}$  for the  description of the Nijmegen
np and pp phase shifts in those channels which
were used in the fit at N$^3$LO, namely S-, P, and D-waves and the
mixing angles $\epsilon_1$ and $\epsilon_2$. 
%\begin{table}[t]
%\caption{$\chi^2/{\rm datum}$ for the description of the Nijmegen
%neutron-proton and proton-proton phase shifts \cite{Stoks:1993tb} as described in
%the text. Only those
%channels are included which have been used in the N$^3$LO fits,
%namely the S-, P- and D-waves and the mixing angles $\epsilon_1$ and
%$\epsilon_2$.  
%\label{tab_chi2_1}}
%\smallskip
%\begin{tabular*}{\textwidth}{@{\extracolsep{\fill}}ccccccccc}
%\hline 
%\hline
%\noalign{\smallskip}
% $E_{\rm lab}$ bin &  CD-Bonn \cite{Machleidt:2000ge} & 
%\multicolumn{2}{c}{ --- Idaho N$^3$LO  \cite{Entem:2003ft}  --- } & 
%\multicolumn{5}{c}{ --- improved chiral potentials at N$^3$LO, this work  --- }
%\\
 %(MeV) &  & (500) & (600) & $R=0.8\,$fm & $R=0.9\,$fm  & $R=1.0\,$fm  & $R=1.1\,$fm  &
% $R=1.2\,$fm
%\smallskip
% \\
%\hline \hline
%%\smallskip
%\multicolumn{8}{l}{neutron-proton phase shifts} \\ 
%0--100 & 0.6 & 1.7 & 5.2 & 0.8 & 0.7 & 0.6 & 0.7 & 1.4 \\ 
%0--200 & 0.6 & 2.2 & 5.3 & 0.8 & 0.7 & 0.6 & 0.8 & 1.8 \\ 
%0--300 & 0.6 & 3.3 & 6.8 & 2.1 & 1.5 & 1.8 & 4.0 & 10.7 \\ [4pt]
%%\smallskip
%\hline 
%% \smallskip
%\multicolumn{8}{l}{proton-proton phase shifts} \\ 
%0--100 & 0.5 & 1.5$^\star$ & 6.7$^\star$    &  1.8 & 0.8 & 0.5 & 1.2 & 4.6 \\ 
%0--200 & 1.3 & 2.9$^\star$ & 11.7$^\star$  &  2.1 & 0.7 & 0.6 & 2.2 & 8.2 \\ 
%0--300 & 1.3 & 5.9$^\star$ & 30.0$^\star$  & 12.0& 3.2 & 7.0 & 24.5 & 66.8 \\[4pt]
%\hline \hline
%\multicolumn{8}{l}{$^\star$The $^1S_0$ partial wave has not been taken into account as
%explained in the text.} \\ 
%\end{tabular*}
%\end{table}
\begin{table}[t]
\caption{$\chi^2/{\rm datum}$ for the description of the Nijmegen
np and pp phase shifts \cite{Stoks:1993tb} as described in
the text. Only those
channels are included which have been used in the N$^3$LO fits,
namely the S-, P- and D-waves and the mixing angles $\epsilon_1$ and
$\epsilon_2$.  
\label{tab_chi2_1}}
\smallskip
\begin{tabular*}{\textwidth}{@{\extracolsep{\fill}}cccccccccc}
\hline 
\hline
\noalign{\smallskip}
 $E_{\rm lab}$ bin &  CD-Bonn & 
\multicolumn{2}{c}{ --- Idaho N$^3$LO   --- } & N$^3$LO of \cite{Epelbaum:2004fk} &
\multicolumn{5}{c}{ --- improved chiral potentials at N$^3$LO, this work  --- }
\\
 (MeV) &  & (500) & (600) & (550/600) & $R=0.8\,$fm & $R=0.9\,$fm  & $R=1.0\,$fm  & $R=1.1\,$fm  &
 $R=1.2\,$fm
\smallskip
 \\
\hline \hline
%\smallskip
\multicolumn{9}{l}{neutron-proton phase shifts} \\ 
0--100 & 0.6 & 1.7 &   5.2  &   1.9 & 0.8 & 0.7 & 0.6 & 0.7 & 1.4 \\ 
0--200 & 0.6 & 2.2 &   5.3   &   2.1 & 0.8 & 0.7 & 0.6 & 0.8 & 1.8 \\ 
0--300 & 0.6 & 3.3 &   6.8  &     6.0 & 2.1 & 1.5 & 1.8 & 4.0 & 10.7 \\ [4pt]
%\smallskip
\hline 
% \smallskip
\multicolumn{9}{l}{proton-proton phase shifts} \\ 
0--100 & 0.5 & 1.5$^\star$ & 6.7$^\star$    &  8.3   &     1.8 & 0.8 & 0.5 & 1.2 & 4.6 \\ 
0--200 & 1.3 & 2.9$^\star$ & 11.7$^\star$  &  14.7 &       2.1 & 0.7 & 0.6 & 2.2 & 8.2 \\ 
0--300 & 1.3 & 5.9$^\star$ & 30.0$^\star$  &  75.3 &      12.0& 3.2 & 7.0 & 24.5 & 66.8 \\[4pt]
\hline \hline
\multicolumn{9}{l}{$^\star$The $^1S_0$ partial wave has not been taken into account as
explained in the text.} \\ 
\end{tabular*}
\end{table}
As a test of our
approach, we first apply it to the CD-Bonn potential of
Ref.~\cite{Machleidt:2000ge}. The resulting values for $\chi^2 /{\rm
  datum}$ clearly indicate that this potential provides a very good
description of both the np and pp phase shifts 
of the NPWA in the whole energy range. Notice that the CD-Bonn
potential was fitted to a considerably larger database as compared to
the NPWA. For the Idaho N$^3$LO
potentials of Ref.~\cite{Entem:2003ft}, the  $\chi^2 / {\rm
  datum}$ appears to be somewhat higher, especially for the version with the
cutoff $\Lambda = 600\,$MeV.  Notice that we did not include the
pp $^1$S$_0$ phase shift when calculated the  $\chi^2/ {\rm
  datum}$ for the Idaho N$^3$LO potentials. This is because the
authors of Ref.~\cite{Entem:2003ft} employed a more elaborated
treatment of IB corrections as compared to the NPWA. This is
especially important for the splitting between the
pp and np $^1$S$_0$ phase shifts and would
result in a very large $\chi^2 /{\rm  datum}$ if the
pp $^1$S$_0$ phase would be included. It is interesting to
compare these findings with $\chi^2 /{\rm  datum}$ for the reproduction
of the real data. The values quoted in Ref.~\cite{Machleidt:2000ge}
for the CD-Bonn potential, namely $\chi^2 / {\rm datum} =1.02$ for
np and $\chi^2 / {\rm datum} =1.01$ for pp data 
below $350\,$MeV and in Ref.~\cite{Entem:2003ft} for the two versions
of the Idaho potentials, namely $\chi^2 / {\rm datum} =1.1$-$1.3$ for
np and $\chi^2 / {\rm datum} =1.5$-$2.1$ for pp data 
below $290\,$MeV, show clearly the same qualitative trend. On the other
hand, it is clear that $\chi^2 /{\rm  datum}$ employed in our analysis
is a much more sensitive quantity and the values of $\chi^2 /{\rm  datum}
\sim 5$ do still correspond to accurate description of real data. 

The results for $\chi^2 /{\rm  datum}$  for the improved chiral
potential of the present work at different values of the cutoff $R$ are
listed in the last five columns of Table \ref{tab_chi2_1}. Given that
the softest cutoff $R=1.2\,$fm corresponds to  the momentum-space
regulator of $\Lambda \sim 330\,$MeV, the large values of $\chi^2/{\rm
  datum}$ for this cutoff in the whole energy range of $E_{\rm lab} =
0-300\,$MeV simply reflect the feature that the potential is used at
energies beyond its applicability range. The same applies, to a lesser
extent, to the  cutoff $R=1.1\,$fm which corresponds to the
momentum-space cutoff of $\Lambda \sim 360\,$MeV. As
expected, decreasing the value of the coordinate-space cutoff $R$
leads to a better description of the phase shifts. The improvement
stops for the hardest considered cutoff of $R=0.8\,$fm.  
Notice that the corresponding momentum cutoff $\Lambda \sim 500\,$MeV is
considerably larger than the one found in Ref.~\cite{Lepage:1997cs},
where the TPEP was neglected. Our findings thus confirm the importance
of the two-pion exchange, see also Refs.~\cite{Rentmeester:1999vw,Birse:2003nz} for a related
discussion. Altogether, the description of the np and
pp phase shifts based on the improved N$^3$LO interactions
is excellent for energies below $200\,$MeV 
and, for the optimal cutoff choice of $R=0.9\,$fm, even up to $E_{\rm lab} = 300\,$MeV.    
 
It is also interesting to compare the reproduction of the Nijmegen
phase shifts at different orders in the chiral expansion. In Table
\ref{tab_chi2_2} we show the corresponding values of $\chi^2 / {\rm
  datum}$ for the cutoff $R=0.9\,$fm.
\begin{table}[t]
\caption{$\chi^2/{\rm datum}$ for the description of the Nijmegen
np and pp phase shifts \cite{Stoks:1993tb} as described in
the text at different
orders in the chiral expansion for the cutoff $R=0.9\,$fm. Only those
channels are included which have been used in the N$^3$LO fits,
namely the S-, P- and D-waves and the mixing angles $\epsilon_1$ and
$\epsilon_2$.
\label{tab_chi2_2}}
\smallskip
\begin{tabular*}{\textwidth}{@{\extracolsep{\fill}}ccccc}
\hline 
\hline 
\noalign{\smallskip}
 $E_{\rm lab}$ bin &  LO   &  NLO   &  N$^2$LO   &  N$^3$LO  
\smallskip
 \\
\hline 
\hline 
%\smallskip
\multicolumn{5}{l}{neutron-proton phase shifts} \\ 
%0--100 & 360(560) & 31(13) & 4.5(1.8) & 0.6 \\ 
%0--200 & 480(700) & 63(39) & 21(5) & 0.6 \\ [4pt]
0--100 & 360 & 31 & 4.5 & 0.7 \\ 
0--200 & 480 & 63 & 21 & 0.7 \\ [4pt]
%\smallskip
\hline 
% \smallskip
\multicolumn{5}{l}{proton-proton phase shifts} \\ 
%0--100 & 5750(15300) & 102(112) & 15(11) & 0.7 \\ 
%0--200 & 9150(16900) & 560(695) & 130(105) & 0.6 \\ [4pt]
0--100 & 5750 & 102 & 15 & 0.8 \\ 
0--200 & 9150 & 560 & 130 & 0.7 \\ [4pt]
\hline 
\hline 
\end{tabular*}
\end{table}
The observed pattern provides yet another indication that the chiral
expansion for the nuclear force converges well, see also
Fig.~\ref{fig:phases_conv}. It is especially comforting to see the
improvement when going from NLO to N$^2$LO which is entirely due to
the subleading TPEP. Notice that the number of adjustable parameters is
the same at NLO and N$^2$LO.  Last but not least, our results seem to
support the validity of Weinberg's power counting and do not indicate
the need for its modification as suggested e.g.~in Refs.~\cite{Nogga:2005hy,Long:2011xw,Birse:2010fj}.

\section{Cutoff dependence}
\def\theequation{\arabic{section}.\arabic{equation}}
\label{sec:CutDep}

\begin{figure}[tb]
\vskip 1 true cm
\includegraphics[width=\textwidth,keepaspectratio,angle=0,clip]{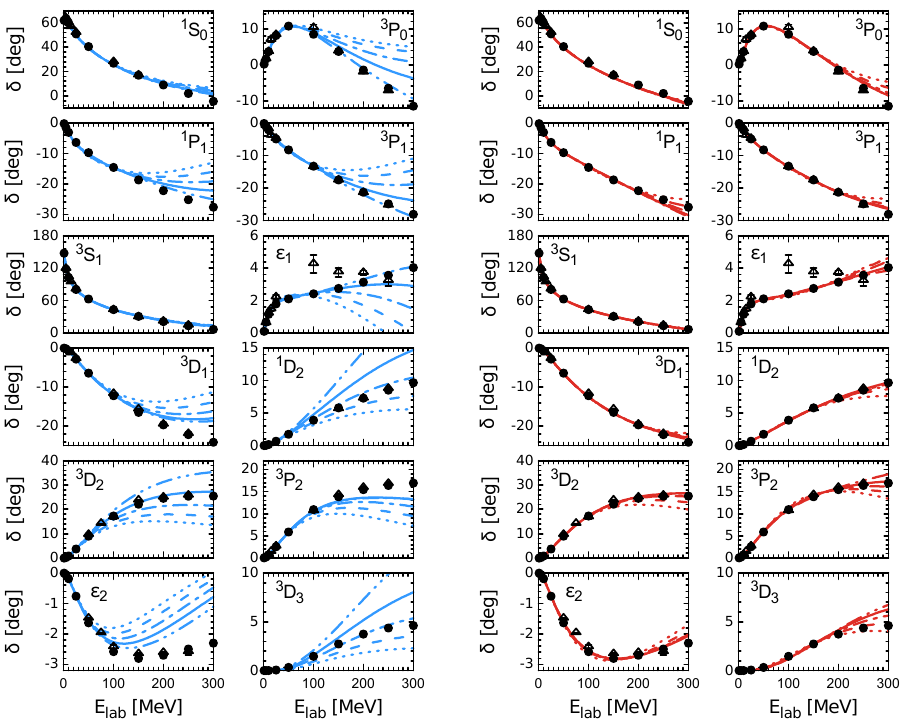}
    \caption{Cutoff dependence of of the phase shifts calculated at N$^2$LO (left panel) and N$^3$LO (right panel). 
Dotted, dashed, dashed-dotted, solid and dashed-double-dotted lines show the results obtained with the cutoffs 
$R=1.2\,$fm, $R=1.1\,$fm, $R=1.0\,$fm, $R=0.9\,$fm and $R=0.8\,$fm,
respectively. 
Only those partial waves are shown which have been used in the fits at N$^3$LO. 
Solid dots and open triangles correspond to the results of the NPWA \cite{Stoks:1993tb}  and the GWU 
      single-energy np partial wave analysis \cite{Arndt:1994br}.
\label{fig:phases_cutoff} 
 }
\end{figure}

We now address in some detail the residual cutoff dependence of our
results. As explained at the beginning of section~\ref{sec:Regular},
the dependence of observables on the cutoff $R$ is \emph{not} completely
removed in our calculations, see however Ref.~\cite{Epelbaum:2012ua} for an
alternative renormalizable approach.  The residual cutoff dependence
can be viewed as an estimation of effects of higher-order contact
interactions beyond the truncation level of the potential, see,
however, the discussion in section \ref{sec:Uncertainty}. One,
therefore, expects the residual cutoff dependence to reduce when going
from LO to NLO/N$^2$LO and  from NLO/N$^2$LO to
N$^3$LO/N$^4$LO. On the other hand, the residual cutoff dependence
at chiral orders  NLO and N$^2$LO as well as N$^3$LO and N$^4$LO is
expected to be of the same size. In Fig.~\ref{fig:phases_cutoff} we
compare the cutoff dependence of the S-, P- and D-wave phase shifts
and the mixing angles $\epsilon_1$ and $\epsilon_2$ at N$^2$LO and
N$^3$LO.   
The 
cutoff dependence at N$^3$LO appears to be very weak in all channels used in the
fit. In particular, it is considerably weaker than the one resulting from our
N$^3$LO potential \cite{Epelbaum:2004fk} where a non-local exponential
regulator was employed for the OPEP, TPEP and the contact interactions.  
The new regularization scheme described in section \ref{sec:Regular} shows
a superior performance at higher energies and produces  only a 
small amount of artefacts (i.e. the residual cutoff dependence) in the
considered energy range.  

\begin{figure}[tb]
\vskip 1 true cm
\includegraphics[width=0.9\textwidth,keepaspectratio,angle=0,clip]{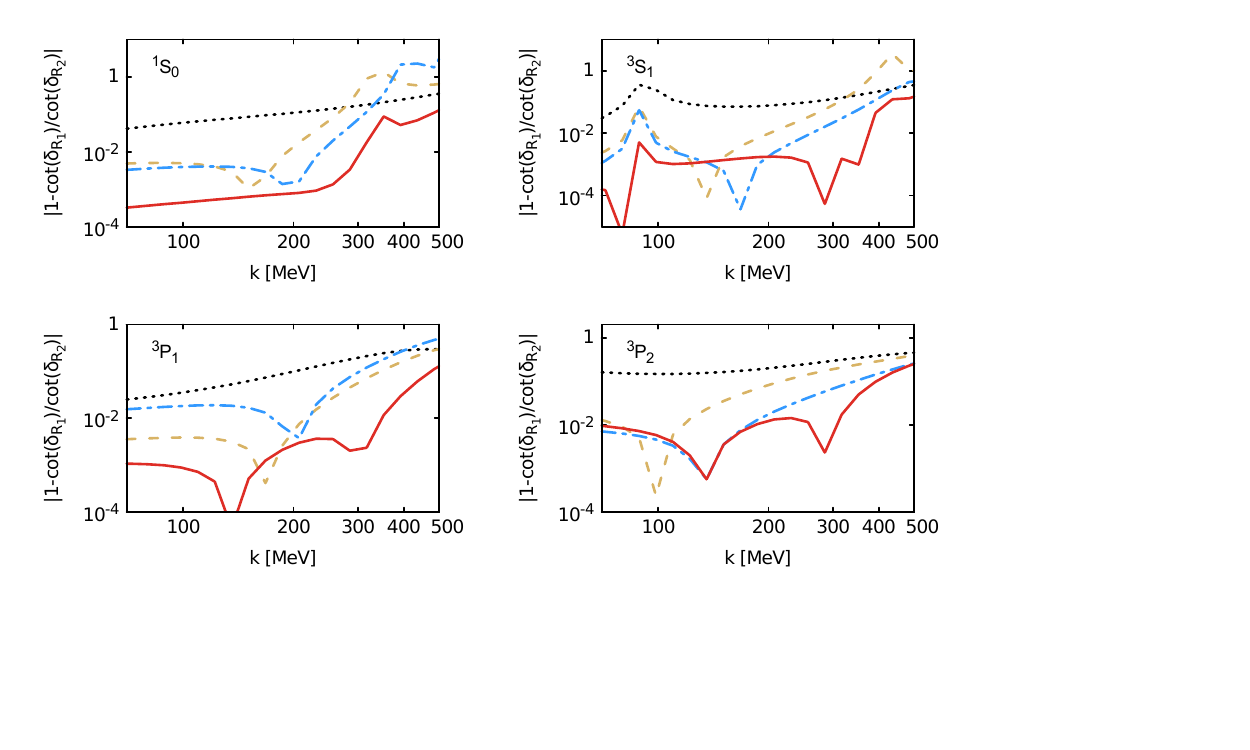}
    \caption{Error plots for np scattering in the $^1$S$_0$,
$^3$S$_1$,  $^3$P$_1$ and  $^3$P$_2$ partial waves as explained in the
text.  Dotted, dashed (color online: brown), dashed-dotted (color
online: blue) and solid (color online: red) lines show the
      results at LO, NLO, N$^2$LO and N$^3$LO, respectively.
\label{fig:grie} 
 }
\end{figure}

To get more insights into the residual cutoff dependence of phase
shift $\delta$  in a given channel, we follow the
lines of Ref.~\cite{Griesshammer:2004pe} and plot in Fig.~\ref{fig:grie} the quantity  
$|  1 - \cot \delta_{R_1}(k) / \cot \delta_{R_2} (k) |  $, where $R_1$
and $R_2$ are two different values of the cutoff, 
as function of the cms momentum $k$. 
Specifically, we choose $R_1=0.9\,$fm and $R_2=1.0\,$fm and restrict ourselves to the np $^1$S$_0$,
$^3$S$_1$,  $^3$P$_1$ and  $^3$P$_2$ partial waves which may serve as
representative examples.  First, the resulting error plots 
demonstrate a very similar cutoff dependence at NLO and N$^2$LO which
is to be expected based on general arguments as discussed above. In
addition,  one
observes that the cutoff dependence reduces significantly in the
whole range of momenta when going from LO to NLO/N$^2$LO and from NLO/N$^2$LO to
N$^3$LO/N$^4$LO. Notice that the appearance of dips in the plots at values of $k$ where the
function $1 - \cot \delta_{R_1}(k) / \cot \delta_{R_2} (k) $ changes
its sign has no significance and should be ignored. Also, the structures seen in the
$^3$S$_1$ partial wave for $k\sim 90\,$MeV and $k\sim 400\,$MeV
($^1$S$_0$ partial wave for $k\sim 350\ldots 400\,$MeV) simply reflect the feature
that $\cot(\pi/2)=0$ ($\cot (0) = \infty$) and should be ignored, too. Concerning the slope of the curves at different
orders, the error plots indicate the appearance of two
different regimes: at low momenta well below the pion mass,  the
slope does not change significantly from order to order and the curves
are nearly horizontal. This is a qualitatively similar pattern to the one reported in
Ref.~\cite{Griesshammer:2004pe}. To understand this feature, we recall
that chiral expansion of the nuclear force is actually a double
expansion in powers of momenta and the pion mass $M_\pi$, see
Eq.~(\ref{ExpPar}). At low momenta, we expect the corrections to
be dominated by powers of $M_\pi/\Lambda_b$ and, therefore, to be nearly independent
on momenta. On the other hand, at momenta above the pion mass, one may expect the
corrections to be dominated by powers of $k/\Lambda_b$.  The increase
of the slope when going from LO to  NLO/N$^2$LO and from NLO/N$^2$LO
to  N$^3$LO can be viewed as a self-consistency check of the
calculation and indicates that the theory is properly renormalized,
see Refs.~\cite{Lepage:1997cs,Furnstahl:2014xsa} for more
details. Finally, we read off from the plots that the breakdown
scale $\Lambda_b$ at N$^3$LO, i.e.~the momenta at which the N$^3$LO
curves cross the ones of lower orders, is about $\sim 500\,$MeV  
for  S-waves and even higher for P-waves. These observations
are in line with our previous findings and, in particular,  with the
size of the LECs accompanying the corresponding contact interactions
which are listed in Table \ref{tab_LEC}.  We will use
$\Lambda_b = 400 \ldots 600\,$MeV, depending on the cutoff $R$, 
in our estimation of the theoretical uncertainties  in section
\ref{sec:Uncertainty}.

\section{Deuteron properties}
\def\theequation{\arabic{section}.\arabic{equation}}
\label{sec:Deuteron}

We now turn to the deuteron properties. First, as already emphasized
in section \ref{sec:PhaseShifts}, we stress that we used the binding
energy $B_d=2.224575\,$MeV \cite{DBD} to constrain the fit. While this choice differs from  our early work
\cite{Epelbaum:2004fk}, it is actually the standard procedure for all high-precision
phenomenological potentials such as the Nijmegen I, II, Reid93,
CD-Bonn and AV18 one. Also the N$^3$LO potential of Ref.~\cite{Entem:2003ft} 
was tuned to reproduce the experimental value of the deuteron binding
energy. We anticipate that relaxing this condition in the fits would
have little impact on few-nucleon observables. 

In Table \ref{tab_deut}, we collect various deuteron properties at
N$^3$LO using different values of the cutoff $R$ in comparison with
the results based on the CD-Bonn \cite{Machleidt:2000ge}, 
N$^3$LO Idaho (500) \cite{Entem:2003ft} and N$^3$LO (550/600) \cite{Epelbaum:2004fk} potentials and with empirical numbers.  
\begin{table}[t]
\caption{Deuteron binding energy $B_d$, asymptotic $S$ state
  normalization $A_S$, asymptotic $D/S$ state ratio $\eta$, radius
  $r_d$ and quadrupole moment $Q$ predicted by various NN potential in
  comparison with empirical information. Also shown is the $D$-state
  probability $P_D$. Notice that $r_d$ and $Q_d$ are calculated
  without taking into account meson-exchange current contributions and
  relativistic corrections.    
\label{tab_deut}}
\smallskip
\begin{tabular*}{\textwidth}{@{\extracolsep{\fill}}llllllllll}
\hline 
\hline
\noalign{\smallskip}
   &  CD-Bonn, & Idaho N$^3$LO & N$^3$LO of \cite{Epelbaum:2004fk} & 
\multicolumn{5}{c}{ --- improved chiral potentials at N$^3$LO, this
  work  --- } & Empirical
\\
   &  \cite{Machleidt:2000ge}  & (500),   \cite{Entem:2003ft} & (550/600) & 
$R=0.8\,$fm & $R=0.9\,$fm  & $R=1.0\,$fm  & $R=1.1\,$fm  &
 $R=1.2\,$fm  &
\smallskip
 \\
\hline \hline
\smallskip
$B_d$ (MeV) & 2.2246$^\star$ & 2.2246$^\star$ & 2.2196   &      2.2246$^\star$ & 2.2246$^\star$ & 2.2246$^\star$ & 2.2246$^\star$ & 2.2246$^\star$ & 2.224575(9) \\ 
$A_S$ (fm$^{-1/2}$) & 0.8846 & 0.8843 &  0.8820  &      0.8843  & 0.8845  & 0.8845  & 0.8846  & 0.8846  & 0.8846(9) \\ 
$\eta$ & 0.0256 & 0.0256 &    0.0254 &           0.0255 & 0.0255 & 0.0256 & 0.0256 & 0.0256 &  0.0256(4)\\ 
$r_d$ (fm) & 1.966 & 1.975 &   1.977  &          1.970 & 1.972 &  1.975 & 1.979 & 1.982 &  1.97535(85)\\ 
$Q$ (fm$^2$)& 0.270 & 0.275 &   0.266 &             0.268 & 0.271 & 0.275 & 0.279 & 0.283 &  0.2859(3)\\ 
$P_D$ ($\%$) & 4.85 & 4.51 &   3.28 &          3.78 & 4.19 & 4.77 & 5.21 & 5.58 &  \\[4pt]
\hline \hline
\multicolumn{9}{l}{$^\star$The deuteron binding energy has been
  taken as input in the fit.} \\ 
\end{tabular*}
\end{table}
In all cases with the exception of the N$^3$LO Idaho potential, the deuteron binding energy is calculated based on
relativistic kinematics, see Eq.~(\ref{kin_np}) and Refs.~\cite{Epelbaum:2004fk,Machleidt:2000ge} for more
details.  We remind the reader that the asymptotic $S$ state
normalization $A_S$  and the asymptotic $D/S$ state ratio $\eta$ are
observable quantities which can be extracted from the S-matrix at the
deuteron pole by means of analytic continuation, see
\cite{Epelbaum:2004fk} and references therein. The empirical
values for these quantities quoted in Table \ref{tab_deut} are taken
from Refs.~\cite{Ericson:1982ei,Rodning:1990zz}. For the deuteron
radius $r_d$, the quoted value corresponds to the so-called deuteron
structure radius which is defined as a square root of the difference
of the deuteron, proton and neutron mean square charge radii and is
taken from Ref.~\cite{Huber:1998zz}.  It agrees well with the earlier
result of 
$r_d = 1.971(5)\,$fm reported  in Ref.~\cite{Martorell:1995zz}.  For
the quadrupole moment $Q$, the experimental value given in Table~\ref{tab_deut} 
is from Ref.~\cite{Bishop:1979zz}. 

Our predictions for $A_S$, $\eta$ and $r_d$ are in excellent agreement
with the empirical numbers. Notice that our calculation of $r_d$ and $Q$ 
does not take into account relativistic and exchange current
contributions. For the radius $r_d$, the corresponding corrections to $r_d^2$ were estimated
in Ref.~\cite{Kohno:1983ru} to give $0.014\,$fm$^2$ while other
calculations quote even smaller numbers. Thus, neglecting these
contributions would affect the
results for $r_d$ at most at the level of $0.2\%$ which is below the
residual cutoff variation $\sim 0.6\%$ for this quantity at N$^3$LO. 

For the quadrupole moment, our predictions underestimate the
experimental value similarly to what is observed for other
modern phenomenological potentials as well as for the chiral N$^3$LO
potentials of Refs.~\cite{Epelbaum:2004fk,Entem:2003ft}. Notice that
the amount of underestimation is largest for the hardest cutoff
$R=0.8\,$fm and reduces strongly for the softest cutoff
$R=1.2\,$fm. We also emphasize that relativistic and meson exchange
current corrections, which are not included in our predictions, were
estimated to increase the value of $Q$ by the amount of $0.010\,$fm$^2$
\cite{Machleidt:2000ge} based on the Bonn one-boson exchange model. This would bring our predictions for the
quadrupole moment in agreement with the experimental value.  
This conclusion is also fully in line with the results of
Ref.~\cite{Phillips:2006im}, where the contributions from the relativistic
corrections and one-pion exchange two-body charge operator were
estimated to be $\Delta Q \simeq 0.008\,$fm$^2$, see also
Ref.~\cite{Piarulli:2012bn} for a related recent work. Adding this correction to
our prediction yields $Q=0.278\ldots 0.291\,$fm$^2$ in agreement with
experiment. The residual cutoff dependence of $Q$ is to be removed by
the leading short-range two-body current. Its required contribution of
the order of $\sim 2\ldots 3\%$ is in agreement with the expected natural
size of the corresponding LEC \cite{Phillips:2006im}. 

It is also instructive to address convergence of the chiral expansion
for the deuteron properties by looking at the predictions at different
orders which are listed in Table \ref{tab_deut2}.  
\begin{table}[t]
\caption{Deuteron properties at various orders in the chiral expansion
  for the cutoff $R=0.9\;$fm in comparison with empirical values. For
  notation see Table \ref{tab_deut}.  
\label{tab_deut2}}
\smallskip
\begin{tabular*}{\textwidth}{@{\extracolsep{\fill}}llllll}
\hline 
\hline
\noalign{\smallskip}
   &  LO & NLO  & N$^2$LO  & N$^3$LO  & Empirical
\smallskip
 \\
\hline \hline
\smallskip
$B_d$ (MeV)  & 2.0235 & 2.1987 & 2.2311 & 2.2246$^\star$ & 2.224575(9) \\ 
$A_S$ (fm$^{-1/2}$) & 0.8333  & 0.8772  & 0.8865  & 0.8845  & 0.8846(9) \\ 
$\eta$ & 0.0212 & 0.0256 & 0.0256 & 0.0255 &  0.0256(4)\\ 
$r_d$ (fm) & 1.990 &  1.968 & 1.966 & 1.972 &  1.97535(85)\\ 
$Q$ (fm$^2$)& 0.230 & 0.273 & 0.270 & 0.271 &  0.2859(3)\\ 
$P_D$ ($\%$) & 2.54 & 4.73 & 4.50 & 4.19 &  \\[4pt]
\hline \hline
\multicolumn{6}{l}{$^\star$The deuteron binding energy has been
  taken as input in the fit.} \\ 
\end{tabular*}
\end{table}
Here we restrict ourselves to a single cutoff choice, namely $R=0.9\,$fm.  One observes a good convergence of the chiral expansion for all
listed quantities with the exception of $P_D$ which is well known to
be not observable.  

\begin{figure}[tb]
\vskip 1 true cm
\includegraphics[width=1.0\textwidth,keepaspectratio,angle=0,clip]{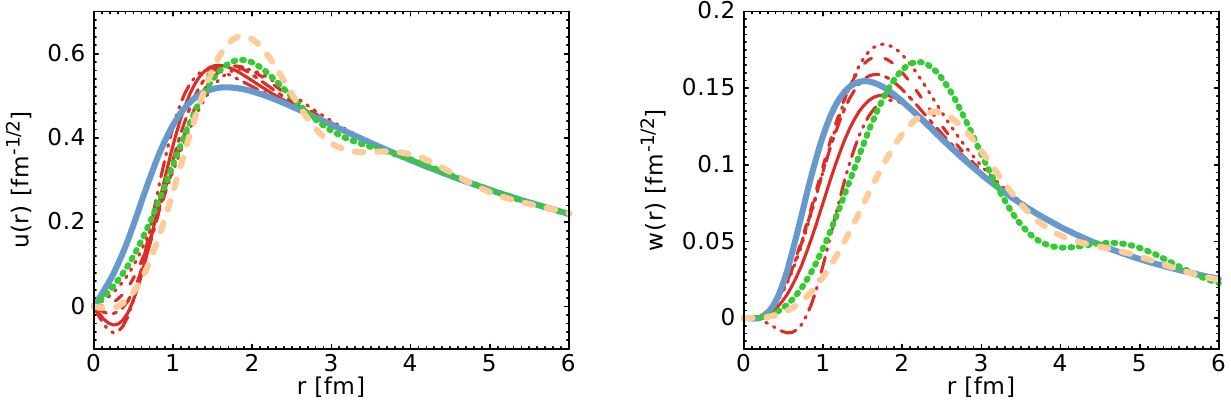}
    \caption{Deuteron wave functions in coordinate space. Thin (color
      online: red) dotted,
      dashed, dashed-dotted, solid and dashed-double-dotted lines show
      the results obtained using the N$^3$LO potentials of this work
      with the cutoffs  $R=1.2\,$fm, $R=1.1\,$fm, $R=1.0\,$fm, $R=0.9\,$fm
      and $R=0.8\,$fm, respectively.  Thick (color online: green)
      dotted, (color online: light brown) dashed and (color online: blue) solid 
      lines refer to  the wave functions of the Idaho (500) N$^3$LO
      potential of Ref.~\cite{Entem:2003ft}, the N$^3$LO (550/600) potential of
      Ref.~\cite{Epelbaum:2004fk} and the CD-Bonn potential \cite{Machleidt:2000ge}.
\label{fig:DWF} 
 }
\end{figure}

Finally, we display in Fig.~\ref{fig:DWF} the deuteron wave functions
calculated using the N$^3$LO potential of the present work in
comparison with those based on the CD-Bonn and the N$^3$LO potentials
of Refs.~\cite{Epelbaum:2004fk,Entem:2003ft}.
As a consequence of the employed regulator, the wave functions based
on the improved chiral potentials are free from the oscillatory
distortions observed in the case of the N$^3$LO potentials of
Refs.~\cite{Epelbaum:2004fk,Entem:2003ft} and are in a very good 
agreement with each other and with the wave functions of the CD-Bonn
potential at distances larger than $r  \sim 2\ldots 3\,$fm.  Notice
that momentum-space deuteron wave functions of the  N$^3$LO potentials of
Refs.~\cite{Epelbaum:2004fk,Entem:2003ft}  show significant 
deviations from the wave functions based on the phenomenological
potentials at $p \sim 400\,$MeV. These deviations were found in Ref.~\cite{Witala:2013ioa} to be
responsible for strong distortions in the predicted shape of the neutron-deuteron
differential cross section around the minimum at the energy of $E_{\rm
  N, \,  lab} = 200\,$MeV. It would be interesting to investigate whether this problem still
persists for the improved chiral potentials. Work along these lines is
in progress.

\section{Estimation of the theoretical uncertainty}
\def\theequation{\arabic{section}.\arabic{equation}}
\label{sec:Uncertainty}

We now turn to the discussion of uncertainty quantification in
nuclear chiral EFT calculations, see Ref.~\cite{Furnstahl:2014xsa} for a recent
paper on this topic. Here and in what follows, our considerations
are restricted to few-nucleon systems, for which the quantum
mechanical $A$-body problem is assumed to be (numerically) exactly
solvable. This certainly applies at least to systems with $A\leq 4$. We will,
therefore, not address uncertainties associated with methods for
calculating observables. 

There are various sources of uncertainties in nuclear Hamiltonian
derived in the framework of chiral EFT which include, see also Ref.~\cite{Furnstahl:2014xsa}:
\begin{enumerate}
\item
\emph{Systematic} uncertainty due to truncation of the chiral
expansion at a given order;
\item
Uncertainty in the knowledge of $\pi N$ LECs which govern the long-range part of the nuclear force;
\item
Uncertainty in the determination of LECs accompanying
contact interactions; 
\item
Uncertainties in the experimental data or, in our case, the NPWA used to
determine the LECs. 
\end{enumerate}
In addition, results of the calculations are expected to show some
sensitivity on the employed regularization framework. This issue will
be addressed at the end of this section. Here and in what follows, we
will primarily concentrate on the first item which, at the curent level of
calculations, we believe to be the dominant source of
uncertainty.  We anticipate, based on the results
reported in
Refs.~\cite{Rentmeester:1999vw,Entem:2002sf,Rentmeester:2003mf,Ekstrom:2014dxa,Perez:2014bua},
that the impact of the uncertainties in $\pi N$ LECs 
and, especially in the order-$Q^2$ ones,  i.e.~the $c_i$, on the
calculated NN observables might be significant. This issue should  be investigated in
a careful and systematic way in the future. A particularly promising
approach to determine the values of the $\pi N$ LECs would be to
perform a simultaneous 
investigation of $\pi N$ scattering and the reaction $\pi N \to
\pi \pi N$, see Ref.~\cite{Fettes:1999wp,Siemens:2014pma} on the chiral EFT treatment of this
process.  Such an analysis goes beyond the scope of this work and is
reserved for a future investigation. 

The uncertainty in the determination of NN contact interaction is clearly
affected by the employed fit procedure such as, in particular,
the choice of energy range and weights adopted in the calculation
of $\chi^2$.  
Following our early work of \cite{Epelbaum:2004fk}, we used
fixed ranges in energies to tune the contact interactions as described
in section~\ref{sec:PhaseShifts} and checked in each case the stability of our
results with respect to their variations. We  did not employ
additional weights  in the $\chi^2$ to account for the
expected increase of the theoretical uncertainty at higher energies,
see Ref.~\cite{Ekstrom:2013kea} for a different approach. 
Both of these issues can, in principle, be addressed in a systematic way within a Bayesian
framework \cite{Schindler:2008fh,Furnstahl:2014xsa}. This topic is
postponed for a future study. Finally, \emph{statistical}
uncertainties for the LECs accompanying NN contact interactions at
N$^2$LO were studied in Ref.~\cite{Ekstrom:2014dxa}. Their impact on
selected pp and np phase shifts can be found in
Table 5 of that work and appears to be negligibly small compared to
the \emph{systematic} theoretical uncertainty to be discussed below.   

Last but not least, there are uncertainties associated with
experimental data or, in our case, with the results of NPWA used as
input in our calculation, see section \ref{sec:PhaseShifts} for more details. 
In particular, we emphasize that the data base used in the NPWA of 1993
has
been largely extended since that time. Specifically, the final
database below $E_{\rm lab}=350\,$MeV used in the NPWA involved 1787 pp
and 2514 np data.  On the other hand, the database used e.g. in the
construction of the CD Bonn potential \cite{Machleidt:2000ge} consists of  2932 pp and
3058 np  data. We expect that increasing the experimental database
should have little 
impact on the resulting phase shifts at the level of the
systematic uncertainty of the NPWA assumed in our work as described in
section \ref{sec:PhaseShifts}. For example, as shown in Table
\ref{tab_chi2_1}, np and pp phase shifts obtained using the CD-Bonn
potential, which is constructed using the extended database, 
are in a very good agreement with the NPWA. Notice, however,
that pp and np phase shifts and mixing angles obtained in the recent coarse-grained potential analysis of NN scattering of 
Ref.~\cite{Perez:2013jpa} do differ significantly from the ones of the CD Bonn
potential and from those of the NPWA and the  Nijmegen
I, II and  Reid 93 potentials.  This is quite surprising given
that this analysis employs essentially the same pp database as the
one used in the construction of the CD-Bonn potential (while the np
database with 3717 data is somewhat larger). Consequently,  using the phase shifts and
mixing angles reported in that work results in fairly large values of
$\chi^2/{\rm datum}$ defined in section \ref{sec:PhaseShifts}, namely 
$\chi^2/{\rm datum} = 4 \ldots 8$. Unfortunately, neither the NPWA \cite{Stoks:1993tb}
nor the coarse grained analysis of Ref.~\cite{Perez:2013jpa} provide
any estimation of the systematic uncertainties so that the
origin and interpretation of these discrepancies remain unclear. 
We do not include the results of Ref.~\cite{Perez:2013jpa} in our
analysis. 

We now address the \emph{systematic} uncertainty of our calculation
due to the truncation of the chiral expansion. To the order we are
working, we expect it to be still the dominant source of the theoretical
uncertainty. Unfortunately, most of the available calculations do not 
address this source of uncertainty or at best
estimate it by means of a residual cutoff dependence, see
e.g.~\cite{Epelbaum:2004fk,Epelbaum:2008ga,Epelbaum:2005pn} and
references therein. Such an approach, however, is well known to suffer from several
deficiencies. First of all, the resulting uncertainty depends on the
employed cutoff range and, therefore, shows some
arbitrariness. Secondly, as already pointed out before, the residual
cutoff dependence measures the contributions due to neglected contact
interactions which appear only at even orders of the momentum
expansion of the NN Hamiltonian. While the residual cutoff dependence
of a given NN observable at LO does indeed measure the size of NLO
corrections, it reflects the sensitivity to the order-$Q^4$ (i.e.~N$^3$LO)
contact interactions at both NLO (order-$Q^2$) and N$^2$LO
(order-$Q^3$). 
For this reason, the uncertainty at NLO and similarly at
N$^3$LO estimated  in this way may be expected to be underestimated. On the other
hand, given that the range of the available momentum-space cutoffs is
rather limited from above both for the conceptual and practical \cite{Machleidt:2009bh} reasons, 
see also the discussion in section
\ref{sec:Regular}, one is forced to employ soft cutoffs in
order to have a cutoff range sufficient for an estimation of the theoretical
uncertainty. Such a procedure is, however, likely to induce large finite-cutoff artefacts
and, therefore,  to unnecessarily overestimate the true theoretical
uncertainty. 

\begin{figure}[tb]
\vskip 1 true cm
\includegraphics[width=1.0\textwidth,keepaspectratio,angle=0,clip]{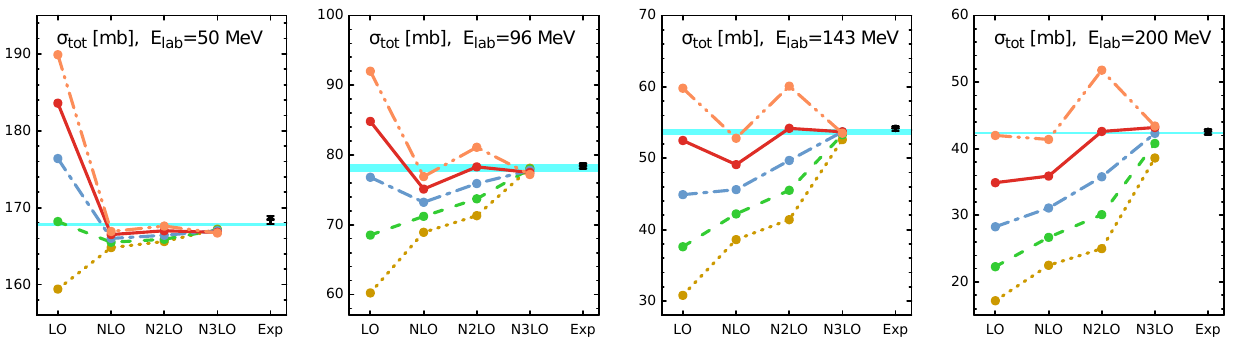}
    \caption{Order-by-order convergence of the chiral expansion for
      the np total cross section at energies of $E_{\rm
        lab}= 50\,$MeV, $E_{\rm
        lab}= 96\,$MeV  and $E_{\rm
        lab}= 143\,$MeV and $E_{\rm
        lab}= 200\,$MeV. Dotted (color online: light brown), dashed
      (color online: green), dashed-dotted (color online: blue),
      solid (color online: red) and dashed-double-dotted (color
      online: pink) lines show the results based on the cutoff $R=1.2\,$fm,
      $R=1.1\,$fm,  $R=1.0\,$fm, $R=0.9\,$fm and $R=0.8\,$fm, respectively.  The
      horizontal band refers to the result of the NPWA with the
      uncertainty estimated by means of deviations from the results
      based on the Nijmegen I, II and Reid 93 potentials as explained
      in the text. Also shown are experimental data of Ref.~\cite{Abfalterer:2001gw}. 
\label{fig:sigmatot}  }
\end{figure}

To illustrate these features, consider the chiral
expansion of the np total cross section at the energies of 
$E_{\rm lab}=50$, $96$, $143$ and $200\,$MeV based on the 
interactions introduced in the previous sections as shown in Fig.~\ref{fig:sigmatot}. 
We also show in this figure by the horizontal band the result of the
NPWA with the assumed theoretical uncertainty and the experimental data of
Ref.~\cite{Abfalterer:2001gw}. 
%Notice that here and in what follows,
%we refrain from showing results based on the hardest employed cutoff of
%$R=0.8$ fm. While this cutoff still yields acceptable results at
%N$^3$LO (with the corresponding potential being highly
%non-perturbative), it is clearly too hard for N$^2$LO where it leads
%to the breakdown of the expansion for D-waves at energies of $E_{\rm
 % lab} =150 \ldots 200$ MeV, see the left panel of
%Fig.~\ref{fig:phases_cutoff}. Given that the description of NN phase
%shifts at N$^3$LO also starts to significantly deteriorate, we
%consider the cutoff $R=0.9$ fm as the optimal choice in the sense that
%it provides the highest possible hard scale and thus the smallest
%expansion parameter  in our calculations.  
The convergence pattern for
the total cross section depicted in Fig.~\ref{fig:sigmatot} shows
the general features one expects to see in chiral EFT: one observes fast convergence at
the lowest energy which becomes increasingly slower at higher
energies. Notice that the large size of
higher-order corrections at the energy of $E_{\rm lab}= 200\,$MeV relative to
the leading ones is actually due to the NLO contributions being
smaller than expected as will be shown below.  
One also observes another feature which persists at all energies,
namely that the size of the N$^2$LO corrections decreases with
increasing the values of $R$. Given that the only new ingredient in
the potential at N$^2$LO is the subleading TPEP, this pattern simply
reflects that the TPEP is stronger cut off for soft cutoff choices. 

The results shown in  Fig.~\ref{fig:sigmatot} provide a good
illustration of the above mentioned issues associated with the 
estimation of the theoretical uncertainty by
means of a cutoff variation.  In particular, 
while the spread in the predictions does, in general,
decrease with the chiral order, it remains nearly the same at NLO and
N$^2$LO. Furthermore, at NLO, it misses (albeit barely) the result of the NPWA 
which is consistent with the expected underestimation of the
theoretical uncertainty at this order. On the other hand, 
while the
spread in the predictions based on different cutoffs is roughly consistent
with the deviations between the theory and the NPWA result 
for the lowest energy, it
appears to significantly overestimate the uncertainty of the
calculation based on lower (i.e.~harder) cutoffs $R$ if one estimates 
it via the deviation between the theory
and the NPWA results. This behavior at high energy suggests that the
spread between the predictions for different values of $R$ is actually
governed by artefacts associated with too soft cutoffs and does
not reflect the true theoretical uncertainty of chiral EFT.  
%Such an analysis therefore does not provide an
%adequate estimation of the theoretical uncertainty in calculations
%based on smaller values of $R$ and, in particular, for the optimal
%choice of $R=0.9$ fm.  
We, therefore, conclude that while being a
useful consistency check of the calculation, cutoff variation in the
employed range does not provide a reliable approach for estimating the
theoretical uncertainty.  As we will show below, 
estimating the uncertainty via the expected size
of higher-order corrections, as it is common e.g.~in the
Goldstone boson and single-baryon sectors of chiral perturbation
theory, provides a natural and more reliable approach which, in
addition, has an advantage to be applicable at any fixed value of the cutoff $R$. 

For a given observable $X( p)$, where $p$ is the cms momentum
corresponding to the considered energy, the expansion parameter in
chiral EFT is given by
\begin{equation}  
\label{expansion}
Q =\max \left( \frac{p}{\Lambda_b}, \; \frac{M_\pi}{\Lambda_b} \right)\,, 
\end{equation}
where $\Lambda_b$ is the breakdown scale. Based on the results
presented in sections \ref{sec:PhaseShifts} and \ref{sec:CutDep}, we
will use $\Lambda_b=600\,$MeV for the cutoffs $R=0.8$, $0.9$ and
$1.0\,$fm, $\Lambda_b = 500\,$MeV for $R=1.1\,$fm and $\Lambda_b=400\,$MeV
for $R=1.2$ to account for the increasing amount of cutoff artefacts
which is reflected by the larger values of $\chi^2/{\rm datum}$ in 
 Table \ref{tab_chi2_1}.  
We have verified the consistency of the
 choice $\Lambda_b=400\,$MeV for the softest cutoff $R=1.2\,$fm by
 making the error plot similar to the one shown in
 Fig.~\ref{fig:grie}.  We can now confront the expected size of
 corrections to the np total cross section at different orders in the
 chiral expansion with the result of the actual calculations. In particular, 
 for the cutoff choice of $R=0.9\,$fm, we obtain 
\begin{eqnarray}
\label{conf_hard}
\sigma_{\rm tot} ( 50\mbox{ MeV}) &=& 183.6_{\rm Q^0} - 17.1_{Q^2 \,
  (\sim 12)} + 0.5_{Q^3 \, (\sim 3) } - 0.2_{Q^4 \, (\sim 0.8)}  =
166.8  \mbox{ mb}\,, \nonumber \\[4pt]
\sigma_{\rm tot} ( 96\mbox{ MeV}) &=& 84.8_{\rm Q^0} - 9.7_{Q^2 \,
  (\sim 11)} + 3.2_{Q^3 \, (\sim 4) } - 0.8_{Q^4 \, (\sim 1.3)}  =
77.5  \mbox{ mb}\,, \nonumber \\[4pt]
\sigma_{\rm tot} ( 143\mbox{ MeV}) &=& 52.5_{\rm Q^0} - 3.4_{Q^2 \,
  (\sim 10)} + 5.1_{Q^3 \, (\sim 4) } - 0.5_{Q^4 \, (\sim 1.8)}  =
53.7  \mbox{ mb}\,, \nonumber \\[4pt]
\sigma_{\rm tot} ( 200\mbox{ MeV}) &=& 34.9_{\rm Q^0} + 1.0_{Q^2 \,
  (\sim 9)} + 6.7_{Q^3 \, (\sim 5) } + 0.6_{Q^4 \, (\sim 2.4)}  =
43.2  \mbox{ mb}\,, 
\end{eqnarray}
see also Fig.~\ref{fig:sigmatot}, while for the softest cutoff $R=1.2\,$fm we find
\begin{eqnarray}
\label{conv_soft}
\sigma_{\rm tot} ( 50\mbox{ MeV}) &=& 159.4_{\rm Q^0} + 5.4_{Q^2 \,
  (\sim 23)} + 0.8_{Q^3 \, (\sim 9) } + 1.6_{Q^4 \, (\sim 3)}  =
167.2  \mbox{ mb}\,, \nonumber \\[4pt]
\sigma_{\rm tot} ( 96\mbox{ MeV}) &=& 60.2_{\rm Q^0} + 8.7_{Q^2 \,
  (\sim 17)} + 2.4_{Q^3 \, (\sim 9) } + 6.8_{Q^4 \, (\sim 5)}  =
78.1  \mbox{ mb}\,, \nonumber \\[4pt]
\sigma_{\rm tot} ( 143\mbox{ MeV}) &=& 30.8_{\rm Q^0} - 7.8_{Q^2 \,
  (\sim 13)} + 2.8_{Q^3 \, (\sim 8) } + 11.2_{Q^4 \, (\sim 5)}  =
52.6  \mbox{ mb}\,, \nonumber \\[4pt]
\sigma_{\rm tot} ( 200\mbox{ MeV}) &=& 17.2_{\rm Q^0} + 5.3_{Q^2 \,
  (\sim 10)} + 2.5_{Q^3 \, (\sim 8) } + 13.6_{Q^4 \, (\sim 6)}  =
38.6  \mbox{ mb}\,.  
\end{eqnarray}
The expected size of NLO, N$^2$LO and N$^3$LO corrections indicated in
the subscripts is estimated as $(p/\Lambda_b)^2$, $(p/\Lambda_b)^3$
and $(p/\Lambda_b)^4$ times the LO result in each particular case.  
The cms momenta corresponding to the energies of $E_{\rm lab} = 50$,
$96$, $143$ and $200\,$MeV are  $p= 153\,$MeV,  $p= 212\,$MeV,  $p= 259\,$MeV
and  $p= 307\,$MeV, respectively. Generally, the estimated size of
corrections at various orders appears to be in a reasonable agreement with their
actual size.  The N$^3$LO corrections are smaller than expected for
$R=0.9\,$fm but turn out to be large for the cutoff  $R=1.2\,$fm at higher
energies. We emphasize that it might be too optimistic to expect a convergent 
expansion at the energies of $E_{\rm lab}= 143$ and $200\,$MeV for the
softest cutoff since the expansion parameter $Q$ in these cases is
larger than $0.5$.  Also the fact that the LO contribution  at the highest
energy for  $R=1.2\,$fm amounts to  less than half of the total result
suggests that this cutoff is not applicable at such an energy. We also
observe an interesting feature that the EFT expansion actually 
converges faster than expected at low energy when soft cutoffs are
employed, see the first line in Eq.~(\ref{conv_soft}) and the left
plot in Fig.~\ref{fig:sigmatot}.  This behavior becomes even more
pronounced at lower
energies. In fact, when reducing the cutoff $R$, we actually
continuously 
integrate out pion physics, and the resulting theory would gradually turn
into pionless EFT if we would further soften the cutoff. At very low
energies with momenta well below the pion mass, pionless EFT, which
correspond to the expansion in $p/M_\pi$, may actually be more
efficient than the expansion in chiral EFT which is controlled by the
parameter $M_\pi/\Lambda_b$. 

Having tested our estimation for the breakdown scale $\Lambda_b$ in
the results for the np total cross section at various chiral orders,
we are now in the position to estimate the theoretical uncertainty of 
our results at N$^3$LO. To be on a conservative side, we will ascribe
the uncertainty $\Delta X^{\rm N^3LO} (p)$ of our N$^3$LO prediction $X^{\rm N^3LO}(p)$ for an observable $X(p)$ via
\begin{eqnarray}
\label{def_error}
\Delta X^{\rm N^3LO} (p) &=& \max  \bigg( Q^5 \times \Big| X^{\rm
    LO}(p) \Big|, \;\;\;  Q^3 \times \Big|
  X^{\rm LO}(p) -   X^{\rm NLO}(p) \Big|, \;\; \;  Q^2 \times \Big|
  X^{\rm NLO}(p) -   X^{\rm N^2LO}(p) \Big|, \nonumber \\
&& {}  \;\;\;\;\;\; \; \; \; \;  Q \times \Big|
  X^{\rm N^2LO}(p) -   X^{\rm N^3LO}(p) \Big|  \bigg)\,,
\end{eqnarray}
where the expansion parameter $Q$ is given by Eq.~(\ref{expansion}) and   
the scale $\Lambda_b$ is chosen dependent of the cutoff $R$ as
discussed above. We emphasize that such a simple estimation of the
theoretical uncertainty does not provide a statistical
interpretation. This can be improved e.g.~by employing a Bayesian
framework \cite{Schindler:2008fh,Furnstahl:2014xsa} and performing
marginalization over higher-order corrections. We postpone such an
analysis for a future study and will adopt the simplified treatment
introduced above here and in what follows. We will further impose  
an additional constraint for the theoretical
uncertainties at NLO and N$^2$LO by requiring them to have at least
the size of the actual higher-order contributions. We emphasize that 
the above way of estimating the uncertainty does not rely on
cutoff variation and can be carried out for any given value of $R$.  

\begin{figure}[tb]
\vskip 1 true cm
\includegraphics[width=1.0\textwidth,keepaspectratio,angle=0,clip]{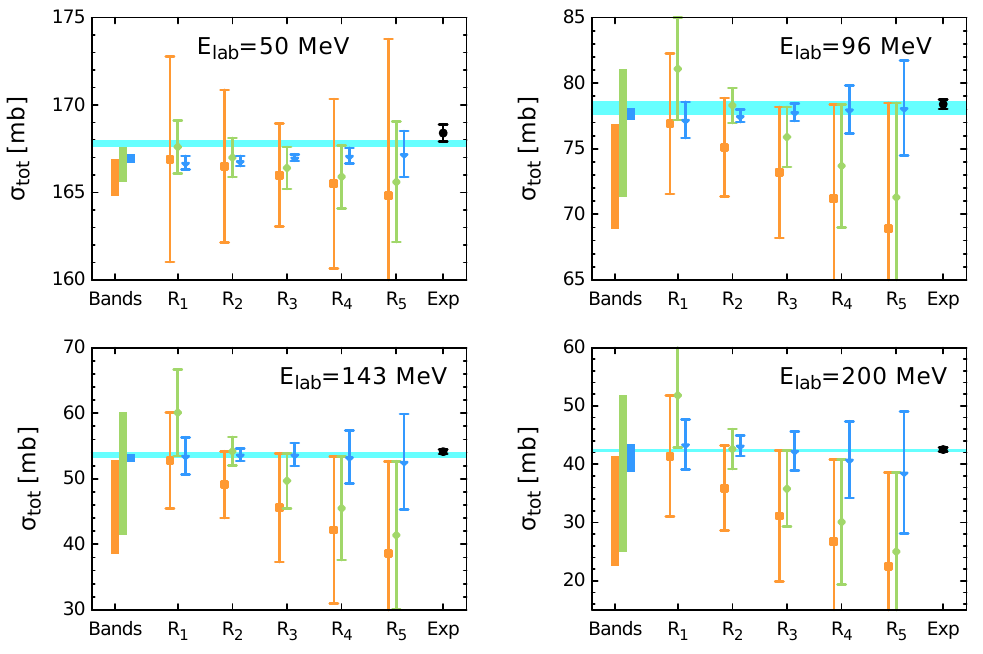}
    \caption{Predictions for the np total cross section based on the improved chiral NN potentials at 
                 NLO (filled squares, color online: orange), N$^2$LO
                 (solid diamonds, color online: green) and
                 N$^3$LO (filled triangles, color online: blue) at the energies of $E_{\rm lab}
                 = 50\,$MeV,  $E_{\rm lab} = 96\,$MeV,  $E_{\rm lab} =
                 143\,$MeV and $E_{\rm lab} = 200\,$MeV  for the different choices of the cutoff: $R_1=0.8\,$fm,    
                 $R_2=0.9\,$fm,  $R_3=1.0\,$fm,  $R_4=1.1\,$fm,  $R_5=1.2\,$fm. Vertical boxes depict the 
                 cutoff dependence of the theoretical predictions at different orders. The
      horizontal band refers to the result of the NPWA with the
      uncertainty estimated by means of deviations from the results
      based on the Nijmegen I, II and Reid 93 potentials as explained
      in the text.   Also shown are experimental data of  Ref.~\cite{Abfalterer:2001gw}. 
\label{fig:sigmatot_error}  }
\end{figure}

Our results for the np total cross section at various orders in the
chiral expansion and for various choices of the cutoff $R$ are shown
in Fig.~\ref{fig:sigmatot_error}. 
Notice that at the smallest energy, we observe deviations between our N$^3$LO
results and the NPWA which are likely caused by the employed
treatment of IB corrections in the $^1$S$_0$ partial way. In
particular, we chose to determine the LECs $C_{1S0}$, $D_{1S0}^1$ and
$D_{1S0}^2$ solely from the pp phase shift and adjusted $\tilde
C_{1S0}^{\rm np}$ to reproduce the np scattering length. The
splitting between the np and pp $^1$S$_0$ phase shifts thus  
comes out as a prediction. It is therefore not surprising that  the  
results for the np $^1$S$_0$ phase shifts show some deviations from
the NPWA. These deviations are expected to be largely reduced at
next-higher order in the chiral expansion. 

In all cases shown in Fig.~\ref{fig:sigmatot_error}, the predicted results calculated using different values
of the cutoff $R$ agree with each other within the theoretical
uncertainty. It is comforting to see that our procedure for estimating
the uncertainty yields the pattern which is qualitatively similar to
the one found based on the $\chi^2/{\rm datum}$ for the description of
the Nijmegen np and pp phase shifts as shown in Table \ref{tab_chi2_1}.   
In particular, we see that the most accurate results at the lowest energy
are achieved with the cutoff $R=1.0\,$fm (with the uncertainty for
the $R=0.9\,$fm case being of a comparable size). At higher energies, the
cutoff $R=0.9\,$fm clearly provides the most accurate choice. 
We also observe that at the lowest energy, the cutoff variation does 
considerably underestimate the theoretical uncertainty at NLO and, to
a lesser extent, at 
N$^3$LO as expected based on the arguments given above. This pattern
changes at higher energies. For example, at $E_{\rm lab}=200\,$MeV, the
cutoff bands at NLO and N$^3$LO appear to be of the same size as the estimated 
uncertainty based on the optimal cutoff $R=0.9\,$fm. It is actually a
combination of two  effects which work against each other
which results in a ``reasonable'' estimation of the NLO and N$^3$LO
uncertainties at higher energies by the cutoff bands: on
the one hand, as already mentioned above, cutoff bands measure the impact of the order-$Q^4$ and
order-$Q^6$ contact interactions and, therefore, underestimate the
uncertainty at NLO and N$^3$LO. On the other hand, at higher energies, 
cutoff bands get increased due to using softer values of $R$ as it is
clearly visible from Fig.~\ref{fig:sigmatot_error}. This conclusion is
further supported by the N$^2$LO cutoff band which strongly
overestimates the estimated uncertainty in the case of $R=0.9\,$fm. 
We also learn from Fig.~\ref{fig:sigmatot_error} that N$^2$LO results
for the total cross section for the 
cutoffs of  $R=0.9\,$fm and $R=1.0\,$fm have the accuracy which is comparable to
N$^3$LO calculations with the softest cutoff $R=1.2\,$fm.  
In summary, we find that the suggested approach for error estimation
is more reliable than the standard procedure by means of cutoff bands and,
in addition, has the advantage of being applicable for a fixed value of
$R$. This allows one to avoid the artificial increase of the
theoretical uncertainty due to cutoff artefacts, the issue which is
especially relevant at high energies where the chiral expansion
converges slower. The issue with using the cutoff bands is expected to
become particularly important at
next-to-next-to-next-to-next-to-leading order (N$^4$LO) in the chiral
expansion. In particular, we expect that the residual cutoff dependence at
N$^4$LO will be comparable to that at N$^3$LO, and that it will
significantly overestimate the real N$^4$LO uncertainty at higher
energies in a close analogy to what is observed at N$^2$LO.  
Last but not least, the ability to carry out
independent calculations with quantified uncertainties also provides a
useful consistency check.

\begin{figure}[tb]
\vskip 1 true cm
\includegraphics[width=1.0\textwidth,keepaspectratio,angle=0,clip]{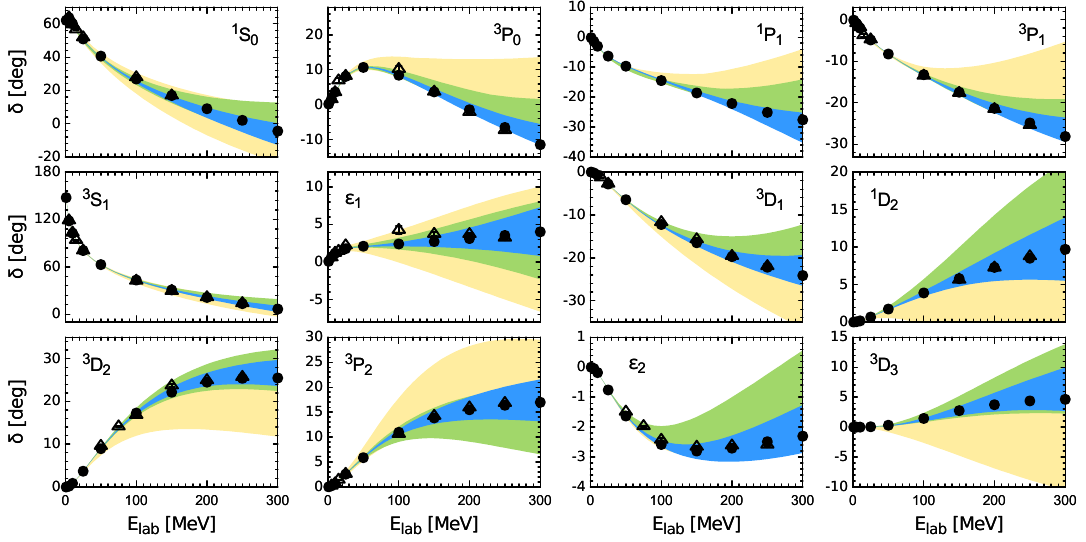}
    \caption{Estimated theoretical uncertainty of the np phase shifts
      at NLO, N$^2$LO and N$^3$LO based on the
      cutoff of $R=0.9\,$fm in comparison with the NPWA \cite{Stoks:1993tb}  (solid dots) and the GWU 
      single-energy np partial wave analysis \cite{Arndt:1994br} (open
      triangles).   The light- (color online: yellow), medium- (color-online: green) and dark- (color-online: blue) 
                  shaded bands depict the estimated theoretical
                  uncertainties at NLO, N$^2$LO and N$^3$LO,
                  as explained in the text. Only those partial waves are shown which 
     have been used in the fits at N$^3$LO. 
\label{fig:phases_errors}  }
\end{figure}

\begin{figure}[tb]
\vskip 1 true cm
\includegraphics[width=1.0\textwidth,keepaspectratio,angle=0,clip]{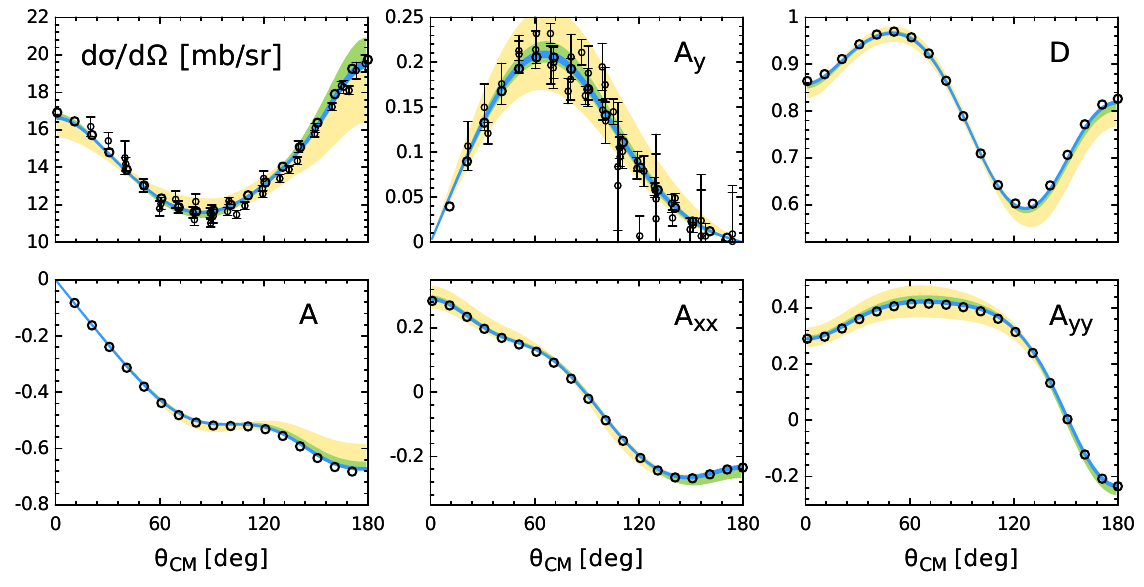}
\vskip 0.2 true cm 
 \includegraphics[width=1.0\textwidth,keepaspectratio,angle=0,clip]{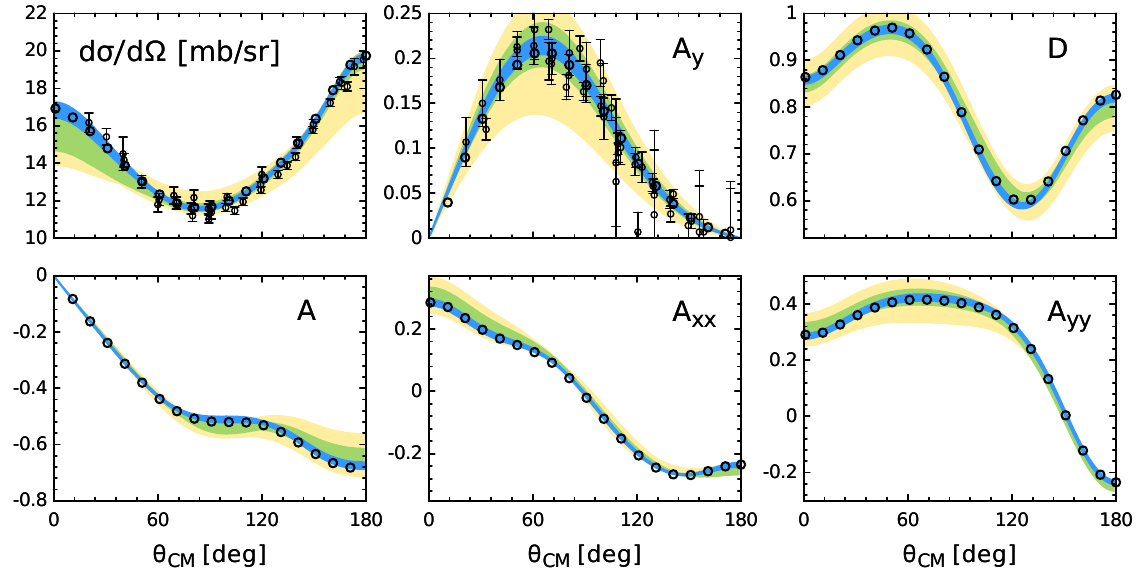}
   \caption{Estimated theoretical uncertainty of the chiral EFT
     results for np differential
     cross section $d \sigma/ d \Omega$, vector analyzing power $A$,
     polarization transfer coefficients $D$ and $A$ and spin
     correlation parameters $A_{xx}$ and $A_{yy}$ at laboratory energy
     of $E_{\rm   lab} = 50\,$MeV. 
                  The light- (color online: yellow), medium- (color-online: green) and dark- (color-online: blue) 
                  shaded bands depict the estimated theoretical
                  uncertainties at NLO, N$^2$LO and N$^3$LO,
                  respectively. Open circles refer to the result of
                  the NPWA. The upper (lower) panel shows the results
                  based on the optimal (softest) cutoff choice of
                  $R=0.9\,$fm ($R=1.2\,$fm).  Data for the cross section are taken from \cite{fink90,mont77} and for the analyzing power
from \cite{fitz80,gar80,lang65,rom78,wil84}. 
\label{fig:NN_25}  }
\end{figure}

\begin{figure}[tb]
\vskip 1 true cm
\includegraphics[width=1.0\textwidth,keepaspectratio,angle=0,clip]{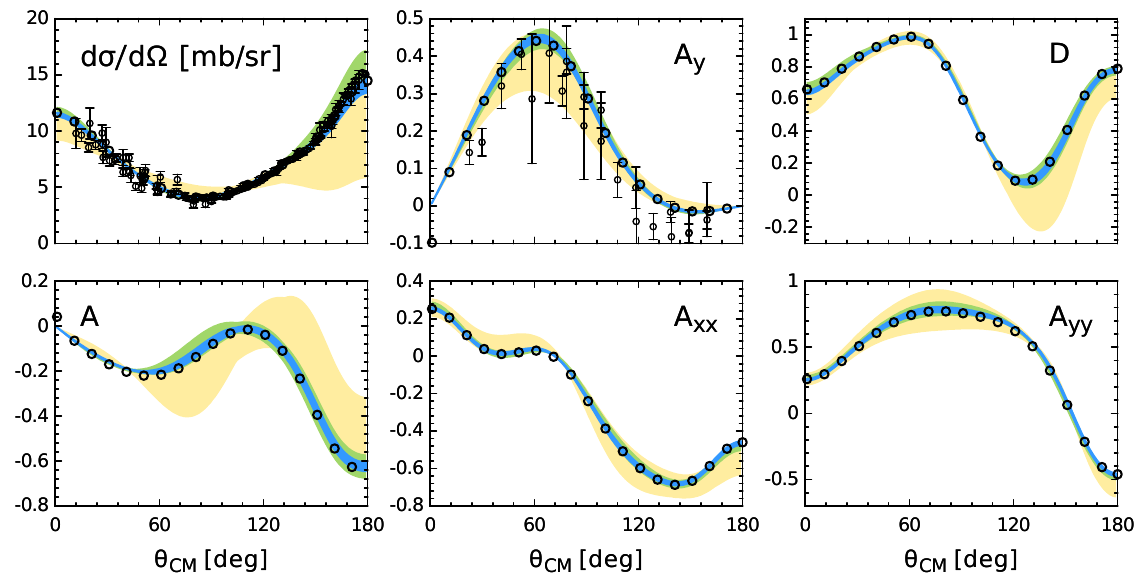}
    \caption{Estimated theoretical uncertainty of the chiral EFT
     results for np differential
     cross section $d \sigma/ d \Omega$, vector analyzing power $A$,
     polarization transfer coefficients $D$ and $A$ and spin
     correlation parameters $A_{xx}$ and $A_{yy}$ at laboratory energy
     of $E_{\rm   lab} = 96\,$MeV calculated using on the
      cutoff of $R=0.9\,$fm. 
                  The light- (color online: yellow), medium- (color-online: green) and dark- (color-online: blue) 
                  shaded bands depict the estimated theoretical
                  uncertainties at NLO, N$^2$LO and N$^3$LO,
                  respectively. Open circles refer to the result of
                  the NPWA. Data for the cross section are taken from \cite{gri58,ra01,roe92}. Data for the analyzing power
are at $E_{\rm lab} = 95\,$MeV and taken from \cite{sta57}. 
\label{fig:NN_96}  }
\end{figure}

\begin{figure}[tb]
\vskip 1 true cm
\includegraphics[width=1.0\textwidth,keepaspectratio,angle=0,clip]{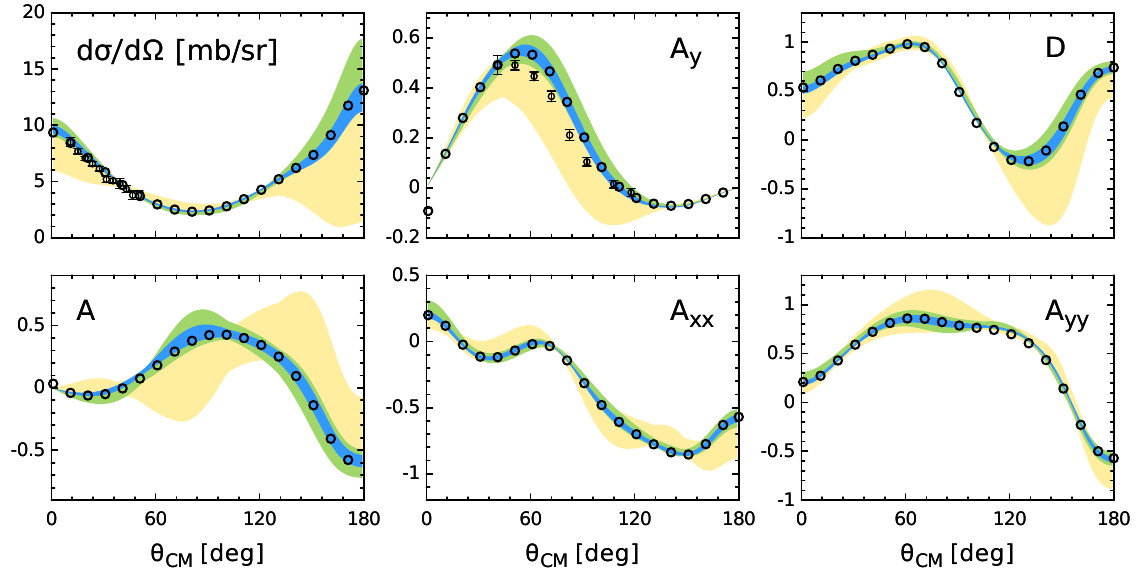}
    \caption{Estimated theoretical uncertainty of the chiral EFT
     results for np differential
     cross section $d \sigma/ d \Omega$, vector analyzing power $A$,
     polarization transfer coefficients $D$ and $A$ and spin
     correlation parameters $A_{xx}$ and $A_{yy}$ at laboratory energy
     of $E_{\rm   lab} = 143\,$MeV calculated using on the
      cutoff of $R=0.9\,$fm. 
                  The light- (color online: yellow), medium- (color-online: green) and dark- (color-online: blue) 
                  shaded bands depict the estimated theoretical
                  uncertainties at NLO, N$^2$LO and N$^3$LO,
                  respectively. Open circles refer to the result of
                  the NPWA. Data for the cross section are at $E_{\rm lab} = 142.8\,$MeV and taken from \cite{ber76} and for the analyzing power
from \cite{kuc61}. 
\label{fig:NN_143}  }
\end{figure}

\begin{figure}[tb]
\vskip 1 true cm
\includegraphics[width=1.0\textwidth,keepaspectratio,angle=0,clip]{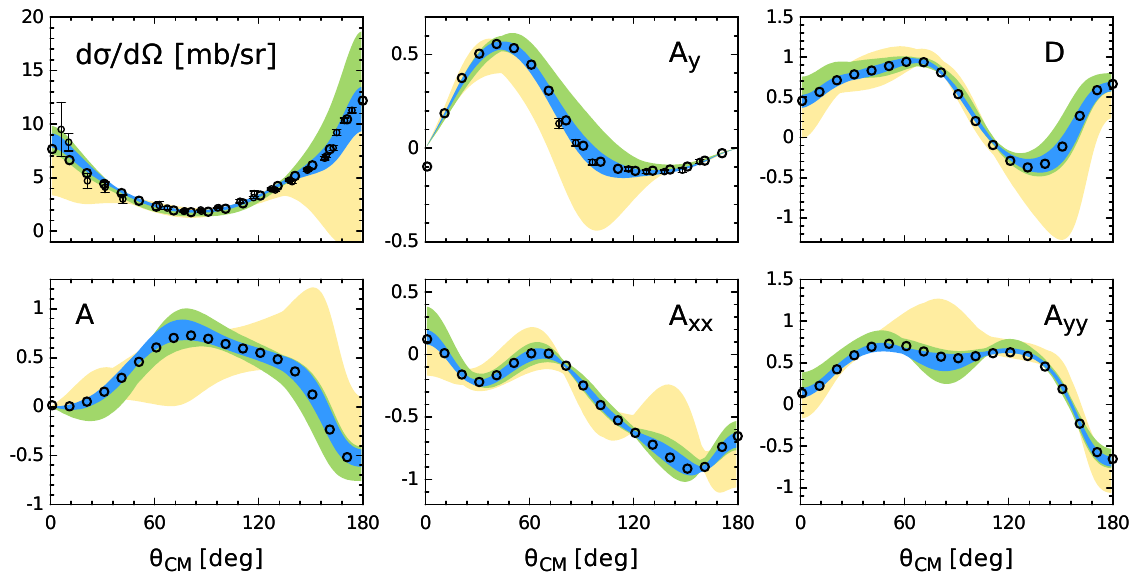}
    \caption{Estimated theoretical uncertainty of the chiral EFT
     results for np differential
     cross section $d \sigma/ d \Omega$, vector analyzing power $A$,
     polarization transfer coefficients $D$ and $A$ and spin
     correlation parameters $A_{xx}$ and $A_{yy}$ at laboratory energy
     of $E_{\rm   lab} = 200\,$MeV calculated using on the
      cutoff of $R=0.9\,$fm. 
                  The light- (color online: yellow), medium- (color-online: green) and dark- (color-online: blue) 
                  shaded bands depict the estimated theoretical
                  uncertainties at NLO, N$^2$LO and N$^3$LO,
                  respectively. Open circles refer to the result of
                  the NPWA. Data for the cross section are at $E_{\rm
                    lab} = 199\,$MeV  from
                  \cite{Thomas:1968zza} and at $E_{\rm
                    lab} = 200\,$MeV from \cite{kaza}. Data for the
                  analyzing power are  at $E_{\rm
                    lab} = 199\,$MeV  from
                  \cite{Thomas:1968zza}.
\label{fig:NN_200}  }
\end{figure}

Next, we show in Fig.~\ref{fig:phases_errors} the estimated
uncertainty of the S-, P- and D-wave phase shifts and the mixing
angles $\epsilon_1$ and $\epsilon_2$ at NLO, N$^2$LO and N$^3$LO based
on $R=0.9\,$fm. 
The various bands result by adding/subtracting the estimated
theoretical uncertainty, $\pm \Delta \delta (E_{\rm lab})$ and $\pm \Delta \epsilon (E_{\rm lab})$, to/from the
results shown in Fig.~\ref{fig:phases_conv}. In a similar way, 
we also looked at selected neutron-proton scattering observables
at different energies shown in Figs.~\ref{fig:NN_25}-\ref{fig:NN_200}.  
For the lowest considered energy of $E_{\rm lab} = 50\,$MeV, we show,
in addition to the results using $R=0.9\,$fm, also our predictions for
the softest cutoff choice of $R=1.2\,$fm.   While the uncertainty is
clearly increased, the results actually still appear to be rather
accurate at this energy. Our results agree with the ones of the NPWA for all
considered observables and energies indicating that the employed way
to estimate the uncertainties is quite reliable. Generally, we find that chiral
EFT at N$^3$LO allows for very accurate results at energies below
$E_{\rm lab} \sim 100\,$MeV and still provides accurate description  
of the data at energies of the order of $E_{\rm lab} \sim 200\,$MeV. These findings are particularly promising for the ongoing studies
of the three-nucleon force whose contributions to nucleon-deuteron
scattering observables are believed to increase at energies above
$E_{\rm N, \, lab} \sim 100\,$MeV. It would be interesting to perform a
similar analysis of nucleon-deuteron scattering data based on the
improved chiral NN potentials in order to see whether accurate
predictions are to be expected at such energies at N$^3$LO. Work along
these lines is in progress.  

Finally, we emphasize that our results depend little on the specific
choice of the regulator function. In order to quantify this
dependence, we performed fits using the cutoff $R=1.0\,$fm but
employing different values of the exponent in Eq.~(\ref{NewReg}), namely
 $n=5$ and $n=7$.  In Table \ref{tab_regul}, we show the resulting
phase shifts in the $^3$S$_1$ and pp $^1$S$_0$, $^3$P$_0$, $^3$P$_1$
and $^3$P$_2$ partial waves at the energies of $10$, $100$ and $200\,$MeV as representative examples. 
\begin{table}[t]
\caption{Selected phase shifts (in degrees) calculated at N$^3$LO using the
  cutoff of $R=1.0\,$fm with the theoretical uncertainty determined
  according to Eq.~(\ref{def_error}) in comparison with the results of the
  NPWA. Also shown are N$^3$LO fits for the same value of $R$ but a
  different functional form of the regulator  with $n=5$ and $n=7$,
  see Eq.~(\ref{NewReg}), and fits based on the spectral function
  regularization with the corresponding cutoff of $\Lambda =1\,$GeV,
  $\Lambda =1.5\,$GeV and  $\Lambda =2\,$GeV.   
\label{tab_regul}}
\smallskip
\begin{tabular*}{\textwidth}{@{\extracolsep{\fill}}lrrrrrrr}
\hline 
\hline
\noalign{\smallskip}
 Lab.~energy &  NPWA \cite{Stoks:1993tb} & our result & DR, $n=5$   & DR, $n=7$  & SFR,
 1.0 GeV &     SFR, 1.5 GeV &      SFR, 2.0 GeV 
 \\[2pt]
\hline \hline
%\smallskip
\multicolumn{8}{l}{proton-proton $^1$S$_0$ phase shift} \\ 
$10$ MeV    & $55.23$ & $55.22 \pm 0.08$&   $55.22$  &   $55.22$ & $55.22$ & $55.22$ & $55.22$  \\ 
$100$ MeV  & $24.99$ & $24.98 \pm 0.60$ &   $24.98$   &   $24.98$ & $24.98$ & $24.98$ & $24.98$ \\ 
$200$ MeV  & $6.55$   & $6.56 \pm 2.2 $ &   $6.55$  &     $6.56$ & $6.56$ & $6.56$ & $6.57$  \\[4pt]
\multicolumn{8}{l}{neutron-proton $^3$S$_1$ phase shift} \\ 
$10$ MeV    & $102.61$ & $102.61 \pm 0.07$ &   $102.61$  &   $102.61$ & $102.61$ & $102.61$ & $102.61$  \\ 
$100$ MeV  & $43.23$ & $43.22 \pm 0.30$ &   $43.28$   &   $43.20$ & $43.17$ & $43.21$ & $43.22$ \\ 
$200$ MeV  & $21.22$ & $21.2 \pm 1.4$ &   $21.2$  &    $21.2$ & $21.2$ & $21.2$ & $21.2$  \\[4pt]
%\smallskip
\multicolumn{8}{l}{proton-proton $^3$P$_0$ phase shift} \\ 
$10$ MeV    & $3.73$ & $3.75\pm 0.04$ &   $3.75$  &   $3.75$ & $3.75$ & $3.75$ & $3.75$  \\ 
$100$ MeV  & $9.45$ & $9.17 \pm 0.30$ &   $9.15$   &   $9.18$ & $9.18$ & $9.17$ & $9.17$ \\ 
$200$ MeV  & $-0.37$ & $-0.1 \pm 2.3$ &  $-0.1$  &     $-0.1$ & $-0.1$ & $-0.1$ & $-0.1$  \\[4pt]
%\smallskip
\multicolumn{8}{l}{proton-proton $^3$P$_1$ phase shift} \\ 
$10$ MeV    & $-2.06$ & $-2.04 \pm 0.01$ &   $-2.04$  &   $-2.04$ & $-2.04$ & $-2.04$ & $-2.04$  \\ 
$100$ MeV  & $-13.26$ & $-13.42\pm 0.17$ &  $-13.43$   &  $-13.41$ & $-13.41$ & $-13.42$ & $-13.42$ \\ 
$200$ MeV  & $-21.25$ & $-21.2 \pm 1.6$ &   $-21.2$  &    $-21.2$ & $-21.2$ & $-21.2$ & $-21.2$  \\[4pt]
%\smallskip
\multicolumn{8}{l}{proton-proton $^3$P$_2$ phase shift} \\ 
$10$ MeV    & $0.65$ & $0.65 \pm 0.01 $ &   $0.66$  &   $0.65$ & $0.65$ & $0.65$ & $0.65$  \\ 
$100$ MeV  & $11.01$ & $11.03 \pm 0.50$ &  $10.97$   &   $11.06$ & $11.07$ & $11.05$ & $11.04$ \\ 
$200$ MeV  & $15.63$ & $15.6 \pm 1.9$ &   $15.6$  &     $15.5$ & $15.5$ & $15.5$ & $15.6$  \\[4pt]
\hline \hline
\end{tabular*}
\end{table}
Clearly, the observed spread in the results is negligibly small
compared to the estimated accuracy of our calculations. Furthermore,
as already pointed out in section \ref{sec:Regular}, the employed
local regularization of the pion-exchange contributions makes the
spectral function regularization obsolet. In particular, phase shifts
resulting from fits using different values of the SFR cutoff
$\Lambda=1\,$GeV, $\Lambda = 1.5\,$GeV and $\Lambda = 2\,$GeV, see  the last three columns in Table
\ref{tab_regul}, are nearly indistinguishable from each other and from
the DR result corresponding to   $\Lambda= \infty$ and shown in the
third column of this Table.

\section{Summary and conclusions}
\def\theequation{\arabic{section}.\arabic{equation}}
\label{sec:Summary}

In this paper we have presented a new generation of NN potentials
derived in chiral EFT up to N$^3$LO. The new chiral forces 
offer a number of substantial improvements as compared to the widely
used N$^3$LO potentials of Refs.~\cite{Epelbaum:2004fk,Entem:2003ft} introduced a decade
ago. First of all, we employ a local regularization scheme for the
pion exchange contributions which, differently to the standard
nonlocal regularization applied e.g.~in Refs.~\cite{Epelbaum:2004fk,Entem:2003ft}, does not 
distort the low-energy analytic structure of the amplitude and, as a
consequence, leads to a better description of phase shifts and
experimental data. The employed regulator, by construction, removes
the short-range part of the chiral two-pion exchange and
thus makes the additional spectral function regularization used in the
potential of Ref.~\cite{Epelbaum:2004fk}
obsolete. This is a particularly welcome feature given that the
expressions for the three-nucleon force at N$^3$LO and N$^4$LO are only available in the
framework of dimensional regularization. 
Further, in contrast to
the earlier studies of  Refs.~\cite{Epelbaum:2004fk,Entem:2003ft}, we have taken all pion-nucleon
LECs and especially the subleading LECs $c_i$ from pion-nucleon scattering without
any fine tuning.  The LECs accompanying NN contact interactions were
determined by fits to the Nijmegen phase shifts and mixing angles for 
five different values of the  coordinate-space
cutoff $R$ chosen in the range of $R=0.8 \ldots 1.2\,$fm and appear to
be of natural size in all
cases. The new N$^3$LO potentials allow for
an excellent description of the Nijmegen np and 
pp phase shifts at energies below $200\,$MeV and, for the cutoff
choices of $R=0.9\,$fm and $R=1.0\,$fm, even up to $E_{\rm lab} =300 \,$MeV. 
Furthermore,  the deuteron properties are accurately described. 
Moreover, the
deuteron wave functions are free from distortions at distances larger
than $r \sim 2 \ldots 3\,$fm which appear for the N$^3$LO potentials
of Refs.~\cite{Epelbaum:2004fk,Entem:2003ft} due to the employed form
of the regulator. We found that the description of the Nijmegen phase shifts
improves substantially when going from LO to NLO, from NLO to N$^2$LO
and from N$^2$LO to N$^3$LO  as one expects for a convergent
expansion. It is worth to emphasize in this connection that the short range part of the NLO and
N$^2$LO potentials involves the same set of operators. Our findings
therefore provide yet another evidence of the subleading two-pion exchange which
was also observed in earlier studies. As an important consistency
check of our approach, we have studied the residual cutoff
dependence of phase shifts at different orders in the chiral
expansion. We found, in particular, that the cutoff dependence is strongly reduced at 
N$^3$LO compared to N$^2$LO in the whole considered range of energies.  

We have also addressed the issue of the uncertainty of our results due to
the truncation of the chiral expansion at a given order. In particular, we
have argued that the standard procedure for error estimation based on
a cutoff variation is not reliable  and proposed a simple alternative
approach by directly estimating the expected size of higher-order
contributions at a given energy. Such a procedure has the advantage of
being applicable for any fixed value of the cutoff so that
calculations based on different
cutoffs  can be used to provide additional consistency
checks. Furthermore, disentangling the error analysis from the cutoff
variation allows one to avoid an unnecessary increase of uncertainty
due to softening the interaction. Notice that the versions of the
potential corresponding to soft choices of the cutoff $R$ may still be
useful for certain kinds of applications
including, in particular, many-body calculations. 
We have applied this
approach to the total np cross section at several energies and have verified
that the results at different chiral orders and for different values
of the cutoff are indeed consistent with each other. We have
furthermore used this method to quantify the theoretical uncertainty
in the description of the np phase shifts as well as  differential cross sections and selected
polarization observables in np scattering. In particular, we found
the N$^3$LO results for np scattering to be very accurate at energies below $\sim 100\,$MeV 
with the corresponding error bands being barely visible and still
rather accurate at the energy of $E_{\rm lab} = 200\,$MeV. In all considered
cases, our results agree with the ones based on the NPWA within the
estimated theoretical accuracy. This gives us additional confidence in
the reliability of the suggested way of quantifying the uncertainty.  
We have furthermore analyzed the uncertainties associated with making a
specific choice of the functional form of the local regulator and
employing the additional spectral function regularization of the
TPEP and found them to be negligible at the level of the estimated
theoretical accuracy at N$^3$LO. 

The improved chiral potentials introduced in this work should provide an
excellent starting point for applications to few-nucleon systems. In
particular, nucleon-deuteron scattering  offers a natural
testing ground for studying the details of the three-nucleon force
which is subject of extensive research \cite{LENPIC}. The existing
calculations based on modern phenomenological potentials suggest that effects of the
three-nucleon force in nucleon-deuteron scattering should be small 
at low energy (except for certain observables like the vector
analyzing power) but become clearly visible at intermediate energies
of $E_{\rm N, \, lab} \sim 70\,$MeV and above. It is encouraging to see
that chiral EFT provides a rather accurate description of NN
scattering in this energy range. We expect a similar
theoretical accuracy for nucleon-deuteron scattering observables, but
this needs to be verified via explicit calculations. Work along these
lines is in progress. For applications to medium-mass and heavy
nuclei based on the continuum methods, the potentials typically need
to be softened by using the renormalization group type techniques such as e.g. the similarity
renormalization group approach, in order to make the many-body problem
numerically tractable.  It remains to be seen whether the new NN
potentials, which do have a substantial amount of high-momentum
components due to the employed local regulator, can be softened
sufficiently without inducing a too large amount of many-body
forces. 

In addition to the already mentioned applications to few- and
many-nucleon systems, this work should be extended in various
directions. First, the calculations should be carried out at
next-higher order in the chiral expansion, see Ref.~\cite{Entem:2014msa} for
a recent work along this line. This would, in particular, provide a
nontrivial check for our estimation of uncertainties at N$^3$LO. 
Secondly, isospin-breaking effects and the role of
the three-pion exchange contributions should be studied in
detail. Furthermore, it is important to quantify the uncertainty
associated with the values of the pion-nucleon LECs and investigate
the possibility of constraining them from NN or even few-nucleon
data in a systematic way. Last but not least, it would be desirable to
employ a more elaborate way of estimating the systematic theoretical
uncertainty which would allow for a statistical
interpretation of the errors. Work along these lines is in progress.

%%%%%%%%%%%%%%%%%%%%%%%%%%%%%%%%%%%%%%%%%%%%%%%%%%%%%%%%%%%%%%%%%%%%%%%%%%%%%%%%%
\section*{Acknowledgments}

We are grateful to Dick Furnstahl for careful reading of the
manuscript and for many useful comments and suggestions. 
This work was supported by the European
Community-Research Infrastructure Integrating Activity ``Study of
Strongly Interacting Matter'' (acronym HadronPhysics3,
Grant Agreement n. 283286) under the Seventh Framework Programme of EU,
 the ERC project 259218 NUCLEAREFT and by the DFG and NSFC (CRC 110).

\bigskip

%%%%%%%%%%%%%%%%%%%%%%%%%%%%%%%%%%%%%%%%%%%%%%%%%%%%%%%
\appendix
\def\theequation{\Alph{section}.\arabic{equation}}
\setcounter{equation}{0}
\section{Scattering amplitude in the partial wave basis}
\label{sec:scatt}

Consider two nucleons moving with momenta $\vec p_1$ and $\vec p_2$.
We use  relativistic kinematics for relating the energy $E_{\rm lab}$ of  the two nucleons
in the laboratory system to the square of the nucleon momentum $\vec p$ in the cms
defined by the condition $\vec p_1 + \vec p_2 =0$. 
As explained in section~\ref{sec:ChiExpansion}, the NN potentials
constructed in the present work are to be used in the
Schr\"odinger equation\footnote{For the np system, this
  equation is correct modulo terms which are proportional to $(m_p
  - m_n )^2$ which are beyond the accuracy of the present calculation.} 
\beq
\label{SE}
\left[ \frac{p^2}{m_N}+ V \right] \Psi 
=  \frac{k^2}{m_N} \Psi  \,.
\eeq
where $m_N = m_p$,   $m_N = m_n$ and $m_N = 2 m_p m_n/(m_p + m_n)$ for
the pp, nn and np systems,
respectively. Here and in what follows we use for all momenta the
notation of e.g.~$p \equiv | \vec p \, |$. 
The relation between  $E_{\rm lab}$ and $k^2$ in the above equation is
based on relativistic kinematics and reads:
\begin{itemize}
\item
Proton--proton case:
\beq
\label{kin_pp}
k^2 = \frac{1}{2} m_p E_{\rm lab}\,.
\eeq
\item
Neutron--neutron case:
\beq
\label{kin_nn}
k^2 = \frac{1}{2} m_n E_{\rm lab}\,.
\eeq
\item
Neutron--proton case:
\beq
\label{kin_np}
k^2 = \frac{m_p^2 E_{\rm lab} (E_{\rm lab} + 2 m_n)}{(m_n+m_p)^2 + 2 E_{\rm lab} m_p}\,.
\eeq
\end{itemize}

The Lippmann-Schwinger equation for the off-the-energy shell
$T$-matrix 
corresponding to Eq.~(\ref{SE}) and projected onto states with orbital angular momentum
$l$, total spin $s$ and total angular momentum $j$
has the form  
\beq
\label{LS}
T^{sj}_{l'l} (p',p; k^2) =  V^{sj}_{l'l} (p',p) + \sum_{l''} \,
\int_0^\infty dq \, {q}^2 \,   V^{sj}_{l'l''} (p',q)
\frac{m_N}{k^2- {q}\,^2 +i\eta}  T^{sj}_{l''l} (q,p; k^2)~,
\eeq
with $\eta \to 0^+$.
In the uncoupled case, $l$ is conserved. The partial wave projected
potential $V^{sj}_{ll'} (p',p)$ can be obtained using the formulae
collected in appendix~B of Ref.~\cite{Epelbaum:2004fk}. 
The relation between the  S- and T-matrices is given by 
\beq
\label{Sdef}
S_{l'l}^{s j} (k) = \delta_{l'l} - i \pi 
 k  m_N  T_{l'l}^{s j} (k, k; k^2)~.
\eeq
The phase shifts in the uncoupled cases can be obtained from the
$S$--matrix via
\beq
S_{jj}^{0j} = \exp{ \left( 2 i \delta_{j}^{0j} \right)} \; , \quad 
S_{jj}^{1j} = \exp{ \left( 2 i \delta_{j}^{1j} \right)} \;,
\eeq
where we have used the notation $\delta^{sj}_l$.
Throughout, we use the so--called Stapp parametrization~\cite{stapp}
of the $S$--matrix in the coupled channels ($j>0$) defined as
\beq
S = \left( \begin{array}{cc} S_{j-1 \, j-1}^{1j} &  S_{j-1 \, j+1}^{1j} \\
S_{j+1 \, j-1}^{1j} &  S_{j+1 \, j+1}^{1j} \end{array} \right)
=
\left( \begin{array}{cc} \cos{(2 \epsilon)} \exp{(2 i \delta^{1j}_{j-1})} &
i \sin{(2 \epsilon)} \exp{(i \delta^{1j}_{j-1} +i \delta^{1j}_{j+1})} \\
i \sin{(2 \epsilon)} \exp{(i \delta^{1j}_{j-1} +i \delta^{1j}_{j+1})} &
\cos{(2 \epsilon)} \exp{(2 i \delta^{1j}_{j+1})} \end{array} \right)~,
\eeq
and is related to another frequently used parametrization due to Blatt and Biedenharn   \cite{Bl52}
in terms of $\tilde{\delta}$ and $\tilde{\epsilon}$ via 
the following equations:
\beq
\label{blattb}
\delta_{j-1} + \delta_{j+1} =  \tilde{\delta}_{j-1} + \tilde{\delta}_{j+1}\,, \quad\quad
\sin ( \delta_{j-1} - \delta_{j+1} ) = \frac{\tan ( 2 \epsilon)}{\tan (2 \tilde{\epsilon})}\,,
\quad\quad
\sin (\tilde{\delta}_{j-1} - \tilde{\delta}_{j+1}) = 
\frac{\sin ( 2 \epsilon)}{\sin (2 \tilde{\epsilon})}\,.
\eeq

For pp scattering, the phase shifts considered in the
present work are of the nuclear plus relativistic Coulomb interaction
with respect to relativistic Coulomb wave functions, i.e.~$\delta
_{C1+N}^{C1}$ using the notation of Ref.~\cite{Stoks:1993tb}. We use
the method proposed by Vincent and Phatak \cite{Vincent:1974zz} to
calculate the corresponding phase shifts and mixing angles in momentum
space, see also Refs.~\cite{Walzl:2000cx,Epelbaum:2004fk,Machleidt:2000ge} for a
description of this approach.

\end{document}